\documentclass[fleqn,usenatbib]{mnras}

\usepackage{newtxtext,newtxmath}
\usepackage[T1]{fontenc}

\DeclareRobustCommand{\VAN}[3]{#2}
\let\VANthebibliography\thebibliography
\def\thebibliography{\DeclareRobustCommand{\VAN}[3]{##3}\VANthebibliography}

\usepackage{graphicx}	% Including figure files
\usepackage{amsmath}	% Advanced maths commands
\usepackage{amssymb}	% Extra maths symbols

\newlength{\figwidth}
\usepackage{color}
\usepackage{gensymb}
\usepackage[normalem]{ulem}

\definecolor{darkgreen}{rgb}{0.13, 0.55, 0.13}

\title[Distribution and kinematics of $^{26}\mathrm{Al}$]{Distribution and kinematics of $^{26}\mathrm{Al}$ in the Galactic disc}

\author[Y. Fujimoto et al.]{
Yusuke Fujimoto,$^{1}$\thanks{E-mail: yfujimoto@carnegiescience.edu}
Mark R. Krumholz,$^{2,3}$
and Shu-ichiro Inutsuka$^{4}$
\\
$^{1}$Earth and Planets Laboratory, Carnegie Institution for Science, 5241 Broad Branch Road, NW, Washington, DC 20015, USA\\
$^{2}$Research School of Astronomy and Astrophysics, Australian National University, Canberra 2611, A.C.T., Australia\\
$^{3}$ARC Centre of Excellence for Astronomy in Three Dimensions (ASTRO-3D), Canberra 2611, A.C.T., Australia\\
$^{4}$Department of Physics, Nagoya University, Furo-cho, Chikusa-ku, Nagoya, Aichi 464-8602, Japan
}

\date{Accepted XXX. Received YYY; in original form ZZZ}

\pubyear{2020}

\begin{document}
\label{firstpage}
\pagerange{\pageref{firstpage}--\pageref{lastpage}}
\maketitle

%It should be a single paragraph not more than 250 words (200 words for Letters).
%No references should appear in the abstract.
\begin{abstract}
$^{26}\mathrm{Al}$ is a short-lived radioactive isotope thought to be injected into the interstellar medium (ISM) by massive stellar winds and supernovae. However, all-sky maps of $^{26}\mathrm{Al}$ emission show a distribution with a much larger scale height and faster rotation speed than either massive stars or the cold ISM. We investigate the origin of this discrepancy using an $N$-body+hydrodynamics simulation of a Milky-Way-like galaxy, self-consistently including self-gravity, star formation, stellar feedback, and $^{26}\mathrm{Al}$ production. We find no evidence that the Milky Way's spiral structure explains the $^{26}\mathrm{Al}$ anomaly. Stars and the $^{26}\text{Al}$ bubbles they produce form along spiral arms, but, because our simulation produces material arms that arise spontaneously rather than propagating arms forced by an external potential, star formation occurs at arm centres rather than leading edges. As a result, we find a scale height and rotation speed for $^{26}\mathrm{Al}$ similar to that of the cold ISM. However, we also show that a synthetic $^{26}\text{Al}$ emission map produced for a possible Solar position at the edge of a large $^{26}\mathrm{Al}$ bubble recovers many of the major qualitative features of the observed $^{26}\mathrm{Al}$ sky. This suggests that the observed anomalous $^{26}\mathrm{Al}$ distribution is the product of foreground emission from the $^{26}\mathrm{Al}$ produced by a nearby, recent supernova.
\end{abstract}


%%%%%%%%%%%%%%%%%%%%%%%%%%%%%%%%%%%%%%%%%%%%%%%%%%

%%%%%%%%%%%%%%%%% BODY OF PAPER %%%%%%%%%%%%%%%%%%

\section{Introduction}

$^{26}\mathrm{Al}$ is a radioactive element with a half-life of 0.7 Myr \citep{NorrisEtAl1983} that forms primarily in massive stars, and is distributed into the interstellar medium (ISM) by supernova (SN) explosions and by the Wolf-Rayet winds that precede them \citep[e.g.][]{Adams2010, LugaroOttKereszturi2018}. Due to the short radioactive decay time, it has been used as a chronometer for the formation events of our Solar system; meteorites' primitive components such as calcium-aluminum-rich inclusions (CAIs), which are the oldest solids in the Solar system, or chondrules, which formed $\sim 1$ Myr after CAI formation, contain significant quantities of the daughter product of $^{26}\mathrm{Al}$, indicating that the Solar system at birth inherited $^{26}\mathrm{Al}$ produced by previous generations of massive star formation \citep[e.g.][]{LeePapanastassiouWasserburg1976, CameronTruran1977, Chevalier2000, GounelleEtAl2009, FujimotoKrumholzTachibana2018}. The $^{26}\mathrm{Al}$ incorporated into the protoplanetary disc was a main heating source for the earliest planetesimals and planetary embryos from which terrestrial planets formed, and it also melted ice layers and dehydrated planetesimals via evaporation, a process that is crucial to determining the bulk water fraction and the habitability of the terrestrial planets \citep[e.g.][]{GrimmMcSween1993, ScherstenEtAl2006, SahijpalSoniGupta2007, LichtenbergEtAl2016, MonteuxEtAl2018, LichtenbergEtAl2019}.

We can study the galactic-scale distribution of $^{26}\mathrm{Al}$ using 1809 keV $\gamma$-ray emission line observations, which trace downward nuclear transitions in the excited $^{26}\text{Mg}$ nuclei left behind when $^{26}\mathrm{Al}$ decays. This line can be detected by space-based observatories such as the Imaging Compton Telescope (COMPTEL) aboard the Compton Gamma Ray Observatory (CGRO) satellite \citep{Pluschke2001} and the Spectrometer on INTEGRAL (SPI) aboard the International Gamma-Ray Astrophysics Laboratory (INTEGRAL) satellite \citep{BouchetJourdainRoques2015}. Although their angular resolutions are only a few degrees, the Galactic sky-map of $^{26}\mathrm{Al}$ shows emission associated with nearby identifiable features, such as the Cygnus, Carina, or Sco-Cen star forming regions, as well as the characteristic spiral arm structures of the Miky Way. These correlations confirm the association between $^{26}\mathrm{Al}$ in the Galaxy and sites of recent massive star formation \citep[e.g.][]{DiehlEtAl2006}.

However, the observed properties of the $^{26}\mathrm{Al}$ emission have proven difficult to understand. One challenge is that the scale height of the emission, as inferred from its angular distribution relative to the Galactic plane, is $\sim 800$ pc \citep{Pleintinger2019, WangEtAl2020}. This is more than an order of magnitude larger than the $\sim 50$ pc scale height of either young, massive stars and star clusters \citep[e.g.,][]{Bobylev16a, Anderson19a, Cantat-Gaudin20a} or the molecular gas from which they form \citep{DameHartmannThaddeus2001}. A second mystery is that the reported mean rotation speed of $^{26}\text{Al}$ is $\approx 100-200$ km s$^{-1}$ greater than that of the Galactic disc \citep{KretschmerEtAl2013}. The origin of this difference between the distribution of $^{26}\text{Al}$ and its purported source is unclear, but two major hypotheses have emerged.

One possibility is that the offset is a result of asymmetry due to spiral arms. \citet{KretschmerEtAl2013} and \citet{KrauseEtAl2015} propose that massive star clusters form predominantly at the leading edges of gaseous spiral arms, and that, when stars in these clusters produce SNe and thence superbubbles, the hot gas blows out into the low-density region forward of the arm, expanding to large heights and leading to faster apparent rotation. \citet{Rodgers-LeeEtAl2019} perform hydrodynamic simulations of gas flowing in a fixed spiral potential rotating at a constant pattern speed, with superbubbles placed in the arms, and show that they recover major qualitative features of the observed distribution, including high-speed and large-height $^{26}\text{Al}$ emission.

A second possibility is that the discrepancies are a result of the Sun being inside a recent SN bubble, so that the anomalous emission reflects local structures rather than the overall Galactic distribution of $^{26}\text{Al}$. Consistent with this hypothesis, \citet{Pleintinger2019} produce synthetic observations of the $^{26}\text{Al}$ sky as seen from a range of possible Solar System positions placed in a simulation of $^{26}\text{Al}$ in the Milky Way by \citet{FujimotoKrumholzTachibana2018}. They show that, while for most possible Solar positions the simulated $^{26}\text{Al}$ emission is more narrowly concentrated around the Galactic plane than is observed, the apparent scale heights are in reasonable agreement if one places the Sun within a superbubble that contains fresh $^{26}\text{Al}$. There is direct evidence that the Sun in fact resides in such a region, known as the Local Hot Bubble (LHB), both from diffuse soft X-rays \citep[and references therein]{Liu17a} and from the direct discovery of live $^{60}\mathrm{Fe}$ in deep-sea crusts and Antarctic snow \citep[e.g.,][]{WallnerEtAl2016, BreitschwerdtEtAl2016, KollEtAl2019}. Since $^{60}\text{Fe}$ has a half-life similar to that of $^{26}\text{Al}$, and both are produced in core collapse SNe, this finding lends strong support to the hypothesis that the Sun might be in a region of locally-enhanced $^{26}\text{Al}$.

One significant limitation of both the simulations of \citet{FujimotoKrumholzTachibana2018} and \citet{Rodgers-LeeEtAl2019} -- the only two published to date to address the Galactic distribution of $^{26}\text{Al}$ -- is that neither contains a realistic treatment of spiral arms and their relationship to star formation. \citet{FujimotoKrumholzTachibana2018} self-consistently include gas self-gravity and star formation, but in their simulations, the stellar and dark matter potential is fixed and axisymmetric, so spiral structures form due to instabilities in the gas alone. By contrast, \citet{Rodgers-LeeEtAl2019} do not include either self-gravity or star formation in their simulations, and instead use a prescription to inject superbubbles along their spiral arms. Moreover, in their simulations, the spiral arms are the result of a fixed potential that rotates at a specified pattern speed, appropriate if the arms of the Milky Way are classical density waves that persist for many Galactic rotations \citep{LinShu1964, BertinLin1996}. However, recent work suggests that the Milky Way's spiral arms are actually material arms rather than density waves; in this picture, they are transient structures formed by swing amplification or local instabilities in the combined gas-star fluid, which continually form, merge, and then shear away. Evidence for this view includes $N$-body simulations showing that the arms are propagating and non-stationary \citep[e.g.][]{WadaBabaSaitoh2011, FujiiEtAl2011, GrandKawataCropper2012a, GrandKawataCropper2012b, DOnghiaVogelsbergerHernquist2013, BabaSaitohWada2013, SellwoodCarlberg2014}, together with analysis of stellar velocity and age distributions from \textit{APOGEE} and \textit{Gaia} data \citep{KawataEtAl2014, GrandEtAl2015, BabaEtAl2018, HuntEtAl2018, SellwoodEtAl2019, KounkelCovey2019, KounkelCoveyStassun2020}.

The nature of spiral arms, and the distribution of star formation and SN-driven bubbles with respect to them, is crucial for the question of $^{26}\text{Al}$ kinematics. As discussed in the review by \citet{DobbsBaba2014}, flow patterns are different depending on how spiral arms form. In simulations such as those of \citet{FujimotoKrumholzTachibana2018}, where only the gas participates in spiral structure, the arms are flocculent and feature only relatively weak shocks. In simulations such as those of \citet{Rodgers-LeeEtAl2019} that use a fixed pattern-speed potential, star formation occurs near the leading edge of the arm because the gas experiences a strong shock as it enters the arms \citep{Fujimoto1968, Roberts1969}. However, if the correct model for the Milky Way's spiral pattern is transient swing-amplified structures, as recent work seems to suggest, then the stellar spiral arm co-rotates with the gas at all radii. As a result the gas follows a colliding flow pattern whereby it slowly falls into the arm from both leading and trailing sides without a strong shock. Since neither of the published simulations correctly model these types of arms, however, the implications of this picture for the $^{26}\text{Al}$ distribution are presently unknown.

In this paper, we will study the galactic-scale distributions and kinematics of $^{26}\mathrm{Al}$ produced in massive stars' stellar winds and SNe, performing an $N$-body+hydrodynamics simulation of the Milky Way galaxy, including the multi-phase ISM and multi-form stellar feedback. The simulation newly includes the transient and recurrent material spiral arms induced by the $N$-body stellar dynamics, which were not included in the previous work of \citet{FujimotoKrumholzTachibana2018}. This paper is organised as follows. In Section~\ref{sec: Methods}, we present our numerical model of a Milky-Way-like galaxy. In Section~\ref{sec: Results}, we show that the results of the spatial distributions and kinematics of $^{26}\mathrm{Al}$ are inconsistent with the estimates from the $\gamma$-ray observations. In Section~\ref{sec: Discussions}, we discuss the possibility that our Solar system resides near to large $^{26}\mathrm{Al}$ bubbles or inside them, and that the $\gamma$-ray observations might have been detecting $^{26}\mathrm{Al}$ emission mostly from the foreground local structures, rather than the background Galactic-scale distributions. In Section~\ref{sec: Conclusions}, we summarise our findings.

\section{Methods}
\label{sec: Methods}

\subsection{Initial conditions}
\label{sec: Initial conditions}

\begin{table*}
	\centering
	\caption{Initial condition characteristics}
	\label{tab:Initial condition characteristics}
	\begin{tabular}{lllll}
		\hline
		& Dark Matter Halo & Stellar Disc & Gas Disc & Stellar Bulge \\
		\hline
		Density profile & \citet{NavarroFrenkWhite1997} & Exponential & Exponential & \citet{Hernquist1990} \\
		\hline
		Structural properties & $M_{200} = 1.074 \times 10^{12}\ \mathrm{M_{\sun}}$ & $M_{\mathrm{disc}} = 3.437 \times 10^{10}\ \mathrm{M_{\sun}}$ & $M_{\mathrm{gas}} = 6.445 \times 10^{9}\ \mathrm{M_{\sun}}$ & $M_{\mathrm{bulge}} = 4.297 \times 10^{9}\ \mathrm{M_{\sun}}$ \\
		& $V_{\mathrm{c,200}} = 150\ \mathrm{km\ s^{-1}}$ & $R_{\mathrm{disc}} = 3.432$ kpc & $R_{\mathrm{gas}} = R_{\mathrm{disc}} = 3.432$ kpc & $\mathcal{R}_{\mathrm{bulge/disc}} = 0.105$ \\
		& $R_{200} = 205.5$ kpc & $h_{\mathrm{disc}} = 0.1\ R_{\mathrm{disc}}$ & $h_{\mathrm{gas}} = h_{\mathrm{disc}} = 0.1\ R_{\mathrm{gas}}$ & \\
		& $c = 10$, $\lambda = 0.04$ & & $f_{\mathrm{gas}} = 0.158$ & \\
		\hline
		Resolution parameters & $m_{\mathrm{halo}} = 1.254 \times 10^5\ \mathrm{M_{\sun}}$ & $m_{\mathrm{disc}} = 3.437 \times 10^3\ \mathrm{M_{\sun}}$ & $\Delta x = 20$ pc & $m_{\mathrm{bulge}} = 3.437 \times 10^3\ \mathrm{M_{\sun}}$ \\
		& $N_{\mathrm{halo}} = 10^7$ & $N_{\mathrm{disc}} = 10^7$ & (at highest AMR level) & $N_{\mathrm{bulge}} = 1.25 \times 10^6$ \\
		\hline
	\end{tabular}
\end{table*}

We make use of the initial conditions for an isolated Milky Way-like galactic disc developed in the AGORA High-resolution Galaxy Simulations Comparison Project \citep{AGORA_KimEtAl2014, AGORA_KimEtAl2016}, which were generated using the \textsc{MakeDisk} code \citep[][see also \citealt{YurinSpringel2014}]{SpringelDiMatteoHernquist2005}. The parameters used in our simulation are summarized in Table~\ref{tab:Initial condition characteristics}. Briefly, we describe the galaxy model in the following. We refer readers to the papers mentioned above for more details.

The dark matter and stars are represented using collisionless particles and are initialized by stochastically drawing from analytic distribution functions. Positions of the dark matter halo particles are initialized to follow an NFW profile \citep{NavarroFrenkWhite1997}, with $M_{200} = 1.074 \times 10^{12}\ \mathrm{M_{\sun}}$, $R_{200} = 205.5$ kpc, the halo circular velocity of $V_{\mathrm{c,200}} = 150\ \mathrm{km\ s^{-1}}$, concentration parameter of $c = 10$ and spin parameter of $\lambda = 0.04$. In practice, \textsc{MakeDisk}, however, follows a \citet{Hernquist1990} profile,
\begin{equation}
    \rho_{\mathrm{halo}}(r) = \frac{M_{\mathrm{halo}}}{2\upi} \frac{a}{r(r+a)^3},
\end{equation}
where $M_{\mathrm{halo}}$ is the mass of the dark matter halo, $r$ is the polar radial coordinate, and $a$ is the halo scale length. The \textsc{MakeDisk} code then transforms this into an approximate NFW profile; tests show that the resulting profiles match the target NFW profile closely \citep{SpringelDiMatteoHernquist2005}.
%the input parameters for the NFW profile into particle distributions which should follow the \citet{Hernquist1990} profile. In fact, the output profile closely matches the NFW profile \citep{SpringelDiMatteoHernquist2005}.

Positions of the stellar disc particles are initilized following
\begin{equation}
    \rho_{\mathrm{disc}}(R, z) = \frac{M_{\mathrm{disc}}}{4\upi R_{\mathrm{disc}}^2 h_{\mathrm{disc}}} \exp \left( -\frac{R}{R_{\mathrm{disc}}} \right) \mathrm{sech}^2 \left( \frac{z}{h_{\mathrm{disc}}} \right),
\end{equation}
where $M_{\mathrm{disc}}$ is the mass of the stellar disc, $R_{\mathrm{disc}}$ is the scale radius, $h_{\mathrm{disc}}$ is the scale height, $R = \sqrt{x^2+y^2}$ is the cylindrical radial coordinate, and $z$ is the vertical coordinate. 

Positions of the stellar bulge particles are initilized to follow a \citet{Hernquist1990} profile as,
\begin{equation}
    \rho_{\mathrm{bulge}}(r) = \frac{M_{\mathrm{bulge}}}{2\upi} \frac{b}{r(r+b)^3},
\end{equation}
where $M_{\mathrm{bulge}}$ is the mass of the stellar bulge, $r$ is the polar radial coordinate, and $b$ is the bulge scale length. The bulge-to-disc mass ratio, $\mathcal{R}_{\mathrm{bulge/disc}} = M_{\mathrm{bulge}} / (M_{\mathrm{disc}} + M_{\mathrm{gas}})$, is set to 0.105.

We include $N_{\mathrm{halo}} = 10^7$ dark matter halo particles, $N_{\mathrm{disc}} = 10^7$ stellar disc particles and $N_{\mathrm{bulge}} = 1.25 \times 10^6$ stellar bulge particles. All particles in each population have uniform masses: $m_{\mathrm{halo}} = 1.254 \times 10^5\ \mathrm{M_{\sun}}$ for the halo population and $m_{\mathrm{disc}} = m_{\mathrm{bulge}} = 3.437 \times 10^3\ \mathrm{M_{\sun}}$ for the disc and bulge populations.

The initial gas distributions on the grid structure are initialized following an analytic density profile as
\begin{equation}
    \rho_{\mathrm{gas}}(R, z) = \frac{M_{\mathrm{gas}}}{4\upi R_{\mathrm{gas}}^2 h_{\mathrm{gas}}} \exp \left( -\frac{R}{R_{\mathrm{gas}}} \right) \exp \left( -\frac{|z|}{h_{\mathrm{gas}}} \right),
\end{equation}
where $M_{\mathrm{gas}}$ is the mass of the gas disc, $R_{\mathrm{gas}}$ is the scale radius, and $h_{\mathrm{gas}}$ is the scale height. We consider an initial gas fraction of $f_{\mathrm{gas}} = M_{\mathrm{gas}} / (M_{\mathrm{disc}} + M_{\mathrm{gas}}) = 0.158$ in order to ensure that the gas fraction after $t = $ 450 Myr, when the galactic disc has fully relaxed into a quasi-equilibrium state,
%without large structural change, 
matches the observed Milky Way value of $\approx 0.14$\footnote{The total stellar and gas masses in the Milky Way are estimated as $5 \pm 1 \times 10^{10}\ \mathrm{M_{\sun}}$ \citep{BlandHawthornGerhard2016} and $8 \times 10^{9}\ \mathrm{M_{\sun}}$ \citep{NakanishiSofue2016}, respectively. That gives us a gas mass fraction of 0.138 in the Milky Way.} of the Milky Way \citep{BlandHawthornGerhard2016, NakanishiSofue2016}. The initial gas temperature in the disc is set to $10^4$ K. The mass distribution of all the four components (halo, disc, bulge and gas) sets an initial rotation curve of the gas disc with circular velocity $V_{\mathrm{c, gas}}(R = 8\ \mathrm{kpc}) = 237\ \mathrm{km\ s^{-1}}$, consistent with observations \citep[e.g.][]{ReidEtAl2019}.

In addition to the gas disc, we include a uniform, low-density gas halo that fills the entire simulation box, with initial density of $7.7 \times 10^{-7}\ \mathrm{cm}^{-3}$, zero initial velocity, and initial temperature of $10^6$ K.

We set the initial abundances of $^{26}\mathrm{Al}$ to $10^{-12}$ throughout the simulation box, though this choice has no practical effect since the initial abundances decay rapidly.

\subsection{N-body dynamics and hydrodynamics}
\label{sec: N-body dynamics and hydrodynamics}

Our simulations follow the evolution of a Milky Way-type galaxy using the adaptive mesh refinement (AMR) code \textsc{Enzo} \citep{BryanEtAl2014, Brummel-SmithEtAl2019}. Rather than using an analytic dark matter and stellar static potential as used in \citet{FujimotoKrumholzTachibana2018}, we employ an $N$-body live dark matter halo and stellar disc, allowing the galactic disc to form spiral arms self-consistently via gravitational interactions among dark matter, stars and gas. The particle dynamics are implemented in the \textsc{Enzo} using a standard particle-mesh scheme. Particle positions and velocities are updated according to the local gravitational acceleration using a drift-kick-drift scheme. Self-gravity of the gas is also implemented. 

We select a direct-Eulerian piecewise parabolic method (PPM) for hydrodynamics, along with a Harten-Lax-van Leer with Contact (HLLC) Riemann solver, to follow the motion of the gas. We treat $^{26}\mathrm{Al}$ as a passive scalar that is transported with the gas, and that decays with the half-life of $t_{1/2} = 0.72$ Myr \citep{NorrisEtAl1983}, following,
\begin{equation}
    \rho (t + \Delta t) = \rho (t) \exp\left(- \ln{(2)} \frac{\Delta t}{t_{1/2}}\right),
\end{equation}
where $\rho(t)$ is the $^{26}\mathrm{Al}$ density of the cell at time $t$ and $\Delta t$ is the local time-step. We do not include dust grain physics because the dust grains and gas are well coupled at the spatial scale we resolve in this simulation.

To model the thermal physics of the gas, we make use of the \textsc{Grackle} cooling library \citep{SmithEtAl2017}. We implement optically thin radiative cooling with tabulated cooling rates for primordial gas and metals assuming ionisation equilibrium between cooling and heating, which is pre-computed with \textsc{Cloudy} \citep{FerlandEtAl2013}. The extragalactic UV background model of \citet{HaardtMadau2012} is used as a photo-heating and photo-ionisation source to calculate the tabulated cooling rate. To model the thermal physics in a Milky-Way-like ISM, we also include gas heating via the photoelectric effect, whereby electrons are ejected from dust grains by far-ultraviolet (FUV) photons. We use the approximate Solar neighbourhood heating rate of $8.5 \times 10^{-26}\ \mathrm{erg\ s^{-1}}$ per hydrogen atom uniformly throughout the simulation box. 

\subsection{Grid structure and refinement criteria}
\label{sec: Grid structure and refinement criteria}

We use an AMR grid setup and refinement criteria similar to those used by \citet{GoldbaumEtAl2015, GoldbaumEtAl2016}.
The galaxy is modelled in a 3D simulation box of $(1.31072\ \mathrm{Mpc})^3$ with isolated gravitational boundary conditions and periodic fluid boundaries. The root grid is $64^3$ cells, on top of which we impose another 5 levels of statically refined regions, enclosing volumes that are successively smaller by a factor of $2^3$. As a result, the galactic disc is enclosed with $(40.96\ \mathrm{kpc})^3$ box with the grid size of 640 pc.

In addition to the static refinement, we impose an additional 5 levels of adaptive refinement, producing a minimum cell size of $\Delta x = 20$ pc, with the following refinement criteria. To keep the dark matter halo properly resolved, we refine a cell if the total mass in particles within the cell exceeds $1.72 \times 10^6\ \mathrm{M_{\sun}}$, or approximately 14 halo particles. To keep the gaseous disc resolved, we refine a cell if the total mass in gas within the cell exceeds $2.15 \times 10^4\ \mathrm{M_{\sun}}$. To avoid artificial fragmentation, we refine a cell if the Jeans length, $L_{\mathrm{J}} = c_{\mathrm{s}} \sqrt{\upi / (G \rho)}$, drops below 16 cell widths, comfortably satisfying the \citet{TrueloveEtAl1997} criterion. In addition, to ensure that we resolve stellar feedback, we require that any computational zone containing a star particle is refined to the maximum level. To keep the Jeans length resolved after collapse has reached the maximum refinement level, we employ a pressure floor such that the Jeans length is resolved by at least 4 cells on the maximum refinement level.

\subsection{Star formation and feedback}
\label{sec: Star formation and feedback}

We use a star formation and feedback prescription similar to that used in \citet{FujimotoKrumholzTachibana2018, FujimotoEtAl2019}. We use a number density threshold for star formation of $13\ \mathrm{cm}^{-3}$ for $\Delta x = 20$ pc. The star formation efficiency per free-fall time for gas above the density threshold is set to 0.01 \citep[see the recent review by][and references therein]{Krumholz19a}. We impose a minimum star particle mass of $300\ \mathrm{M_{\sun}}$ and form star particles stochastically rather than spawn particles in every cell at each time-step. In practice, all star particles are created via this stochastic method.
After forming star particles stochastically, the corresponding gas mass is reduced from the star forming cell.

To model SN explosions and substantial ionising luminosities from massive stars, we use the \textsc{Slug} stellar population synthesis code \citep{daSilvaFumagalliKrumholz2012, da-Silva14b, Krumholz15b}; for each star particle, the \textsc{Slug} stochastically draws individual stars from the initial mass function (IMF), tracks their mass- and age-dependent ionising luminosities and stellar wind mechanical luminosities\footnote{Stellar wind mechanical luminosities in \textsc{Slug} follow the prescription given in \citet{Roy20b}.}, determines when individual stars explode as SNe, and calculates the resulting injection of $^{26}\mathrm{Al}$ using the stellar mass-dependent yield table of \citet{SukhboldEtAl2016}. As noted in \citet{FujimotoKrumholzTachibana2018}, we modify the table by doubling the $^{26}\mathrm{Al}$ yield to ensure that the steady-state mass ratio of $^{60}\mathrm{Fe}/^{26}\mathrm{Al}$ is consistent with observations\footnote{There was a typo in the footnote 2 in \citet{FujimotoKrumholzTachibana2018}. The observed steady-state mass ratio of $^{60}\mathrm{Fe}/^{26}\mathrm{Al}$ should be 1.24, not 0.34. This can be calculated by $(A_{^{60}\mathrm{Fe}} / A_{^{26}\mathrm{Al}}) \cdot (F_{^{60}\mathrm{Fe}} / F_{^{26}\mathrm{Al}}) \cdot (\tau_{^{60}\mathrm{Fe}} / \tau_{^{26}\mathrm{Al}})$, where $A$ is the atomic weight, $F$ is the line flux, and $\tau$ is the half life. However, we keep the factors by which we modify the yield table because of consistency between our previous and current works. In addition, the factors required to be corrected are only a few, which would not change our results substantially.}.

Using the quantities calculated by the \textsc{Slug} code, we include stellar feedback from photoionisation, SNe and stellar winds, following the prescription of \citet{GoldbaumEtAl2016}. Briefly, the H~\textsc{ii} region photoionisation feedback is implemented by heating cells within the Str\"{o}mgren radius, which is estimated from the total ionising luminosity from each star particle each time-step calculated by the \textsc{Slug} code, to $10^4$ K. For each SN that occurs in any given time-step, we add a total momentum of $8 \times 10^5\ \mathrm{M_{\sun}\ km\ s^{-1}}$, directed radially outward in the 26 neighbouring cells. The injected momentum we use is a few times higher than the commonly used value of $\sim 3 \times 10^5\ \mathrm{M_{\sun}\ km\ s^{-1}}$ \citep[e.g. ][]{KimOstriker2015}, because a clustering of SNe can enhance the deposited momentum per SN by a factor of 4 \citep{GentryEtAl2017}. The total net increase in kinetic energy in the cells surrounding the SN host cell is then deducted from the available budget of $10^{51}$ erg, and the balance of the energy is then deposited in the SN host cell and its neighbouring cells as thermal energy. The stellar wind feedback is a new feature which was not included in \citet{FujimotoKrumholzTachibana2018, FujimotoEtAl2019}. We include the effect of stellar winds by adding thermal energy, which is calculated by the \textsc{Slug} code, to each star particle's host cell. We include gas mass injection from stellar winds and SNe to each star particle's host cell each time-step. Note that \citet{SukhboldEtAl2016} report only the total amount of $^{26}\text{Al}$ ejected in pre-SN winds by the stars they model, and do not provide a time-resolved estimate of $^{26}\text{Al}$ production in winds. We must therefore adopt a description for how the $^{26}\text{Al}$ return in the pre-SN winds is distributed in time. For this purpose, we distribute the $^{26}\mathrm{Al}$ mass from massive stars to the host cell each time-step, assuming that the cumulative fraction of the total wind yield of $^{26}\mathrm{Al}$ that has been ejected is proportional to the fraction of the total wind mass. In practice this choice makes little difference, because almost all of the mass and all of the $^{26}\text{Al}$ are ejected during the brief Wolf-Rayet period at the end of a star's life, which lasts for much less than even the shortest galactic timescales (e.g., the vertical oscillation period in the disc). Thus, regardless of our choice of time distribution, there is little opportunity for redistribution of the wind-carried $^{26}\text{Al}$ prior to SNe.

\section{Results}
\label{sec: Results}

In this section we primarily focus on the properties of the gas disc at $t = 650$ Myr, which we show below is a point at which our simulated Milky Way analog has settled into statistical steady state. Unless stated otherwise, all figures below refer to the state of the simulation at this time.

\subsection{Evolution of the galactic disc}
\label{sec: Evolution of the galactic disc}

\begin{figure}
    \centering
	\includegraphics[width=\figwidth]{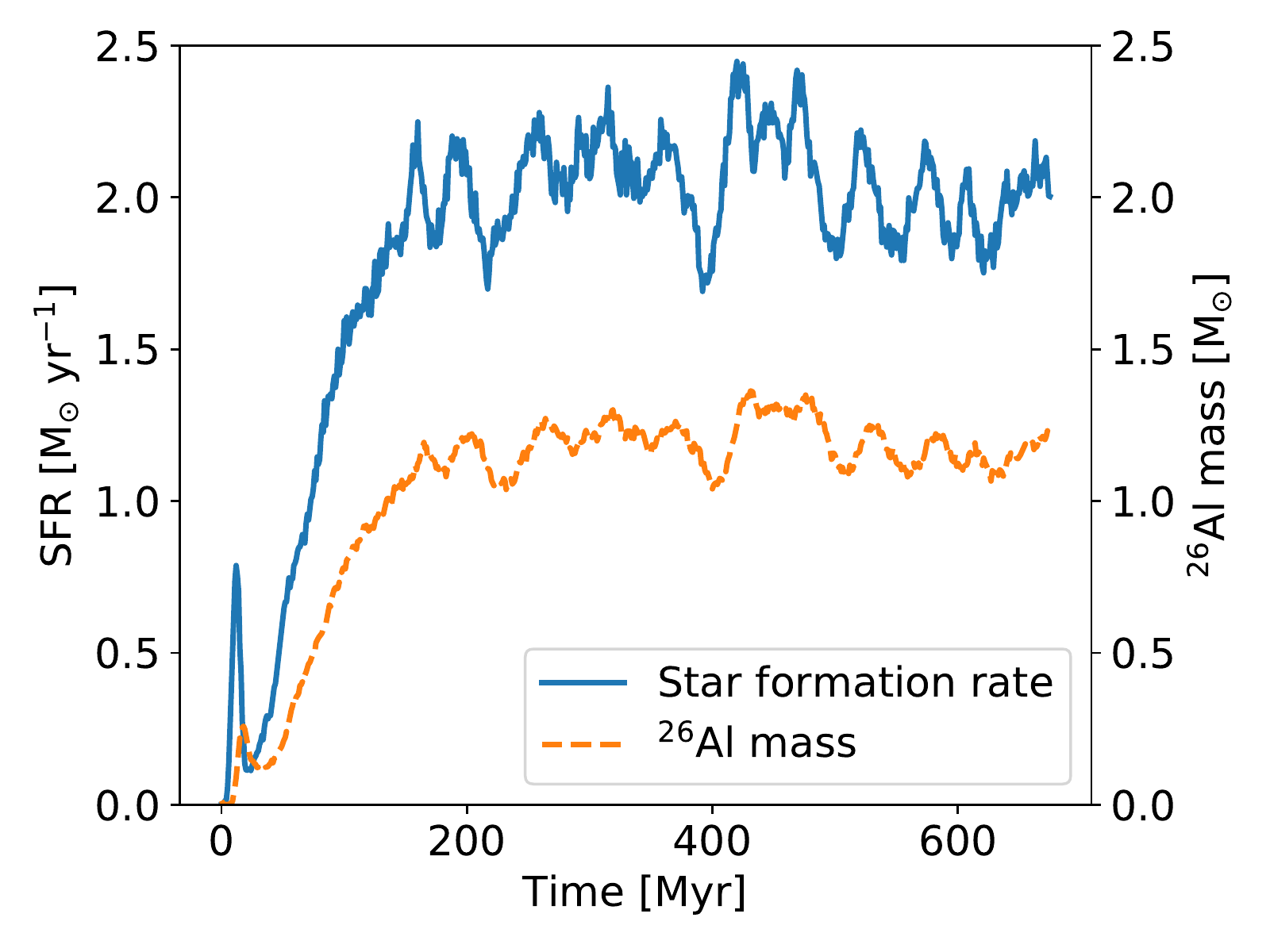}
    \caption{The time evolution of the total star formation rate (SFR: blue solid line) and the total mass of $^{26}\mathrm{Al}$ (orange dashed line) in the galactic disc.}
	\label{fig: time_evolution_sfr_26Al}
\end{figure}

To determine the equilibrium distribution of $^{26}\mathrm{Al}$ in the galactic disc, we run the simulation until the simulated galactic disc settles into a statistical steady state. Fig.~\ref{fig: time_evolution_sfr_26Al} shows time evolution of the total SFR and total $^{26}\mathrm{Al}$ mass within the galaxy. After $t \sim 200$ Myr, which corresponds to one rotation period at 8 kpc from the galactic centre, the galactic disc settles into a steady state, without large changes in the SFR or $^{26}\mathrm{Al}$ mass. In the equilibrium state the SFR is $\sim 2\ \mathrm{M_{\sun}\ yr^{-1}}$, consistent with the observed Milky-Way SFR \citep{MurrayRahman2010, RobitailleWhitney2010, ChomiukPovich2011, LicquiaNewman2015}. We also confirm that the gas mass fraction in the disc to be $\sim 0.14$, which is consistent with observations \citep{BlandHawthornGerhard2016, NakanishiSofue2016}, after $t = 400$ Myr.

\begin{figure*}
    \centering
	\includegraphics[width=0.99\textwidth]{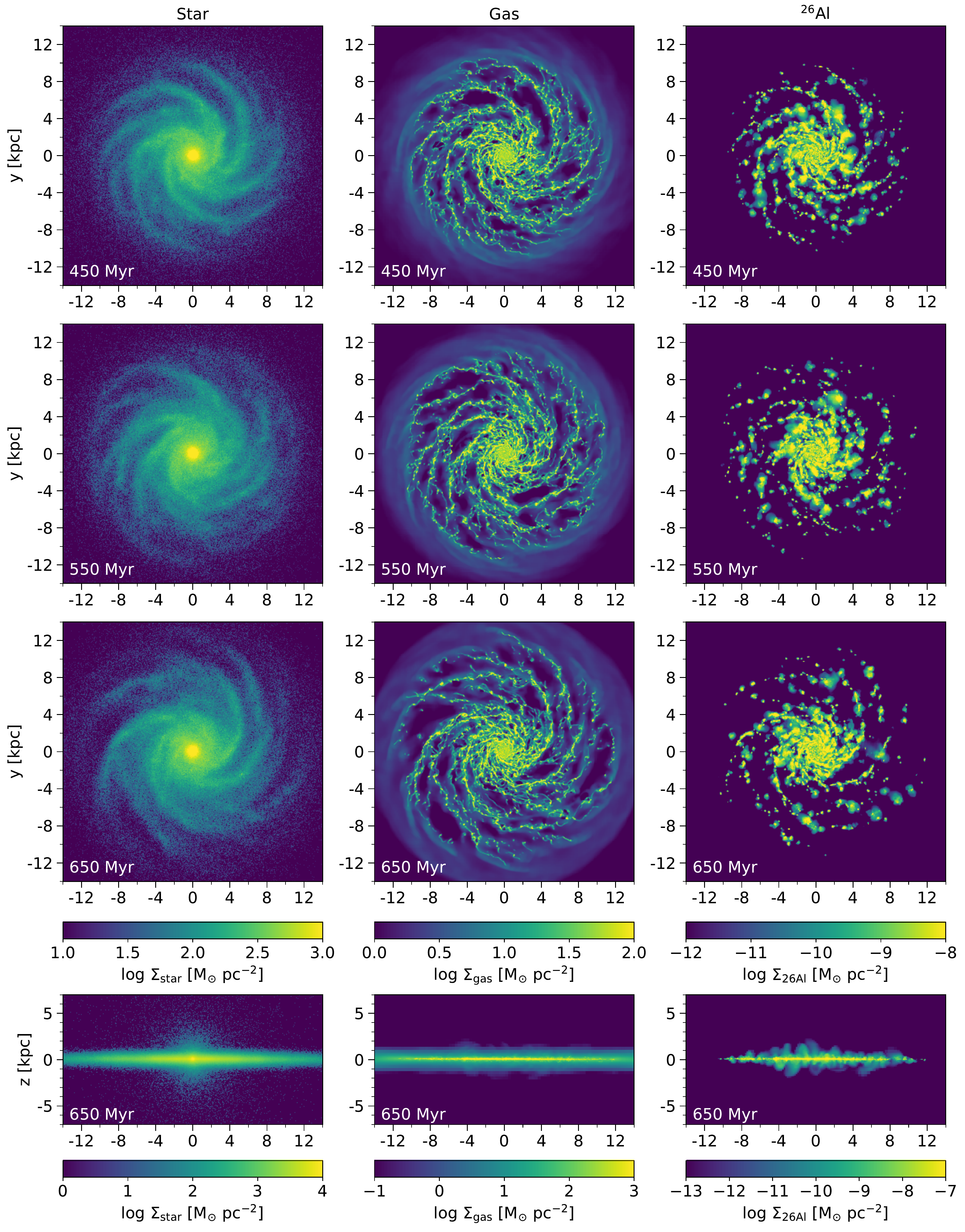}
    \caption{The morphology of the galactic disc. From left to right, panels show the stellar (left), gas (middle), and $^{26}\mathrm{Al}$ surface densities (integrated over $-20 < z < 20$ kpc). From top to bottom, panels show the disc at $t =$ 450, 550 and 650 Myr. The bottom row also shows the disc viewed edge-on at 650 Myr. The galactic disc rotates clockwise.}
    \label{fig: galaxy_projections}
\end{figure*}

\begin{figure*}
    \centering
	\includegraphics[width=\textwidth]{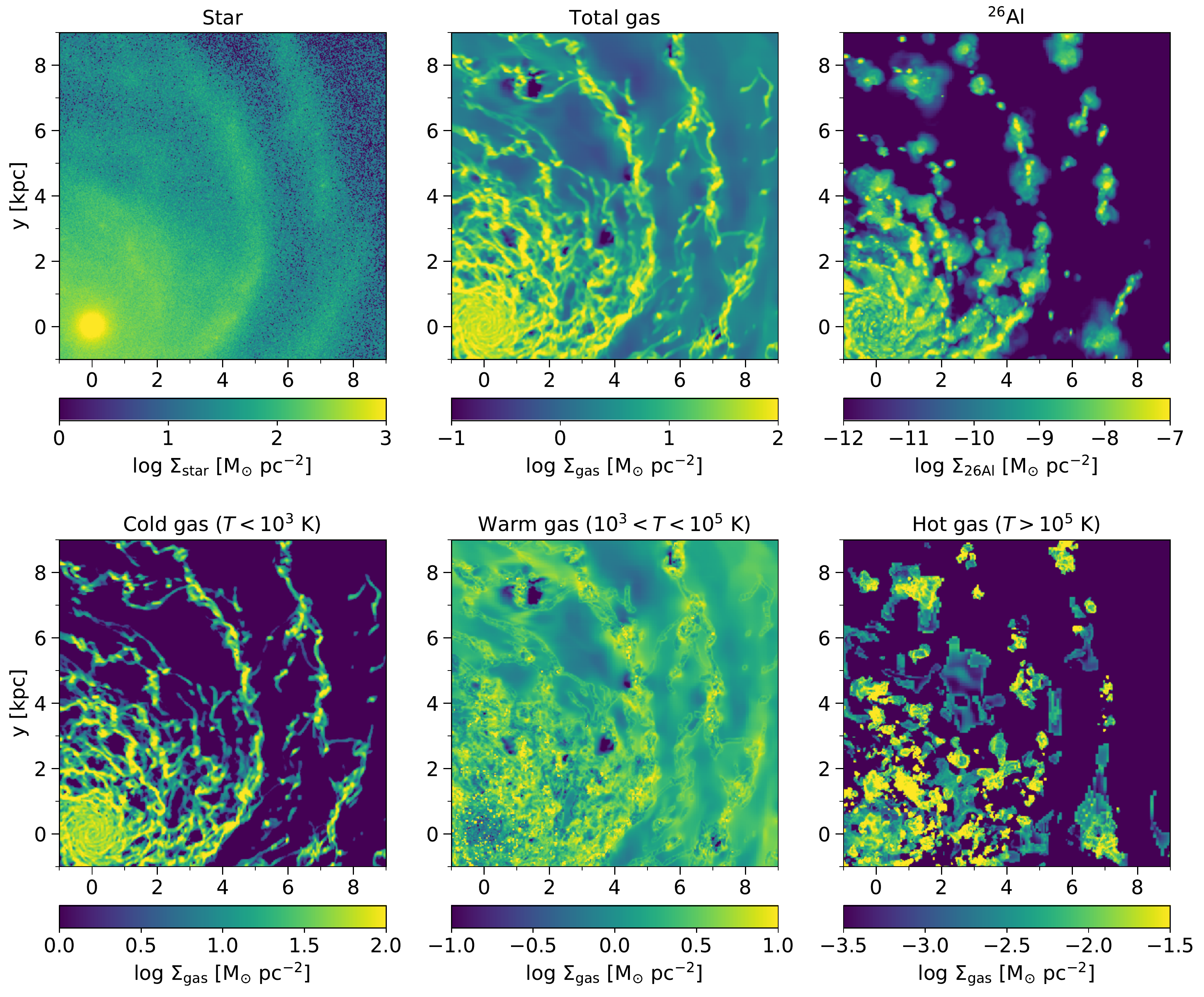}
    \caption{Zoom-in images of the galactic disc at $t =$ 650 Myr. Panels show surface densities (integrated over $-250 < z < 250$ pc) of stars (top left), total gas (top middle), $^{26}\mathrm{Al}$ (top right), cold (bottom left), warm (bottom middle) and hot-phase gas (bottom right).  The galactic centre is located at the lower left corner at $(x, y) = (0, 0)$ kpc, and the galactic disc rotates clockwise.}
    \label{fig: zoom_projections}
\end{figure*}

The morphology of the galactic disc does not drastically change once the disc has been in an equilibrium state. Fig.~\ref{fig: galaxy_projections} shows global distributions of stars, gas and $^{26}\mathrm{Al}$ in the galactic disc at $t = $ 450, 550, and 650 Myr, and
Fig.~\ref{fig: zoom_projections} shows zoom-in images of stars, gas and $^{26}\mathrm{Al}$ at $t = $ 650 Myr, together with distributions of gas separated into three temperature phases: the cold-phase gas ($T < 10^3$ K), the warm-phase gas ($10^3 < T < 10^5$ K), and the hot-phase gas ($T > 10^5$ K).
During this time period, transient and recurrent multi-arm spirals continue to appear due to the $N$-body stellar dynamics and their interactions with the gas, as shown in previous $N$-body+hydrodynamics galaxy simulations \citep[e.g.][]{WadaBabaSaitoh2011, GrandKawataCropper2012a, BabaSaitohWada2013}. The individual spiral arm pattern changes on a time scale of $\sim$ 100 Myr, due to bifurcation or merging of arms. Because $^{26}\mathrm{Al}$ is ejected from young stellar populations, it is found mainly in large bubble structures along the gaseous spiral arms.

\subsection{$^{26}\mathrm{Al}$ distribution in the galactic plane}
\label{sec: 26Al distributions in the galactic plane}

\begin{figure*}
    \centering
	\includegraphics[width=0.7\textwidth]{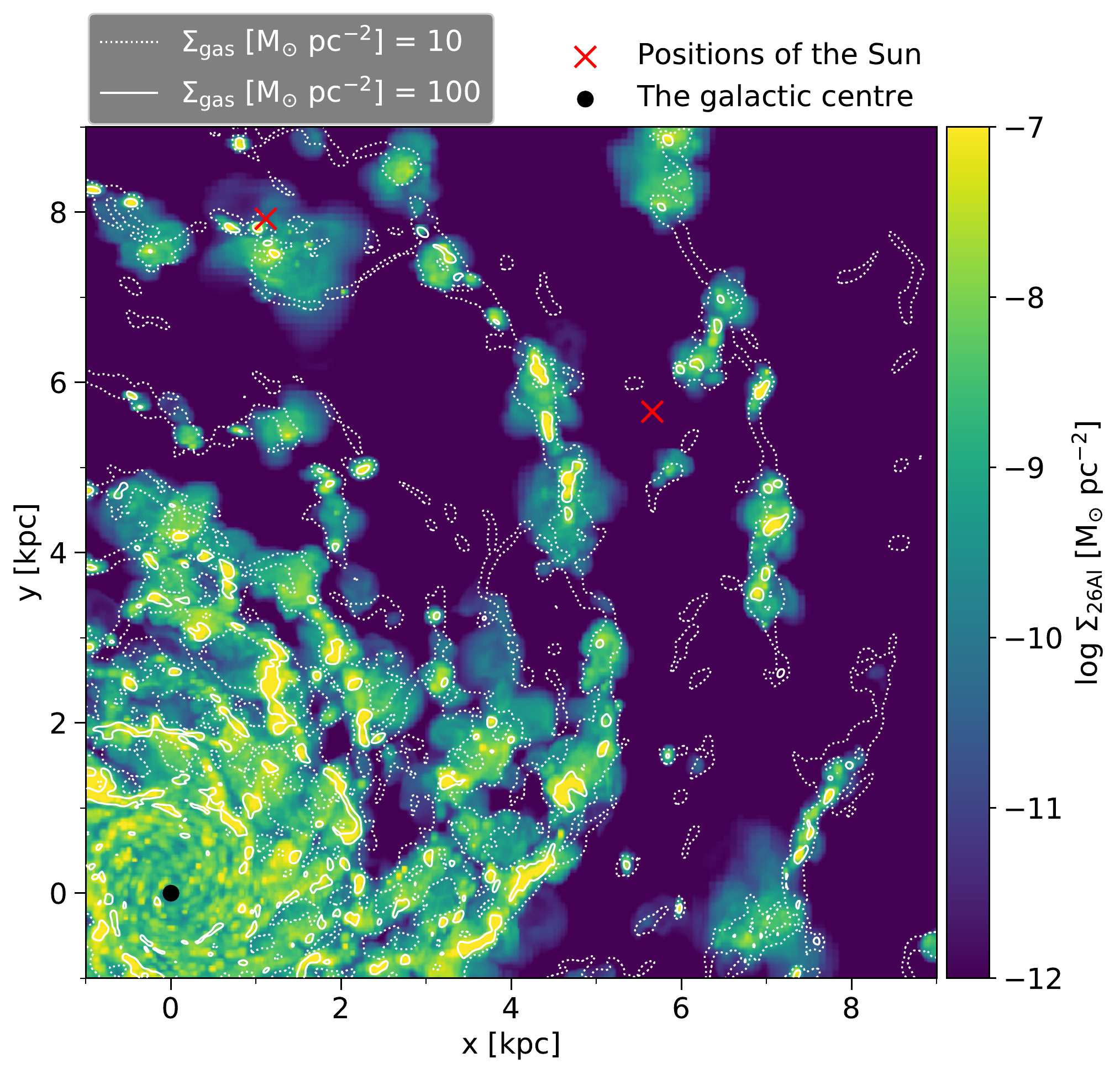}
    \caption{Zoom-in image of the gas (contours) and $^{26}\mathrm{Al}$ (colours) surface densities (integrated over $-250 < z < 250$ pc). The black dot shows the galactic centre, and the galactic disc rotates clockwise. The red x marks show the positions of the Sun (observer) used in the sky maps in Fig.~\ref{fig: longitude profiles}.}
    \label{fig: 26Al_projection_with_gas_contour}
\end{figure*}

Fig.~\ref{fig: 26Al_projection_with_gas_contour} shows a zoom-in image of $^{26}\mathrm{Al}$ distribution in the disc viewed face-on, together with gas distribution overlaid with contours. It shows that sub-kpc scale $^{26}\mathrm{Al}$ bubbles are located along the gaseous spiral arms. However, there is no systematic preference for the direction relative to the spiral arms in which these bubbles extend; they expand to both leading and trailing sides of the spiral arm. This is a different picture from that proposed by \citet{KretschmerEtAl2013} and \citet{KrauseEtAl2015}; they speculate that $^{26}\mathrm{Al}$ bubbles have a preferential expansion towards the leading side of the spiral arms, under the assumption that the arms are density waves. In this scenario, gas compression and star formation occur on the leading edge of the arms where the gas shocks upon entry, and the density gradient through the arm then confines the SN-heated gas and allows expansion only forward of the arm, where the density is lowest. On the other hand, the spiral arms in our simulation are material arms where stars and gas reside throughout the lifetime of the arm, rather than a density wave that propagates through the disc \citep{WadaBabaSaitoh2011}. The gas slowly falls into spiral arms from both leading and trailing sides as a colliding flow, and then massive stars form at the spiral potential minimum. In this scenario there is no preferred direction for the density gradient, and thus SN-produced $^{26}\mathrm{Al}$ bubbles expand to both sides of the spiral arms, both leading and trailing.

\subsection{Vertical distribution of $^{26}\mathrm{Al}$}
\label{sec: Vertical distributions of 26Al}

\begin{figure*}
    \centering
	\includegraphics[width=0.9\figwidth]{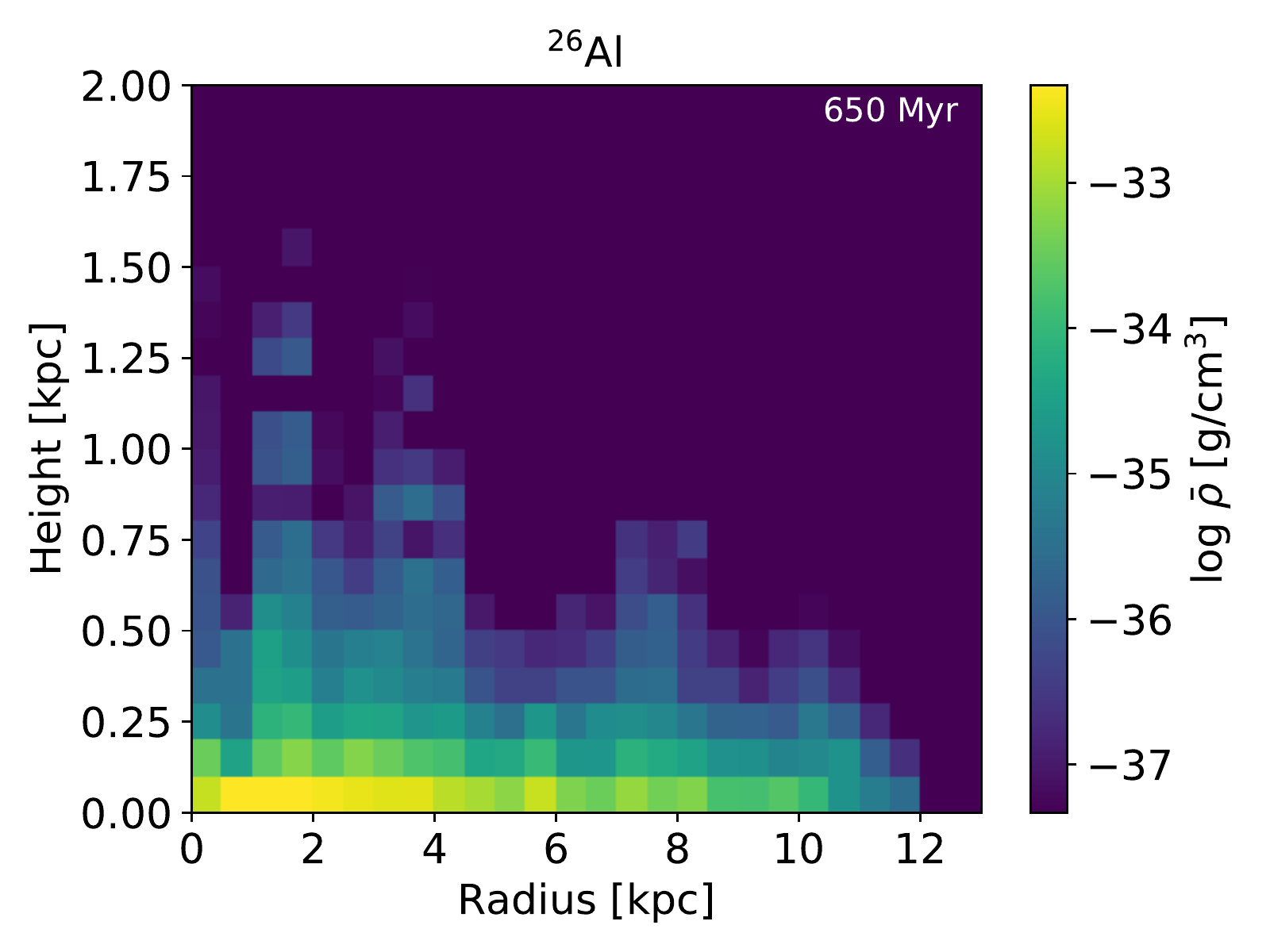}
	\includegraphics[width=0.9\figwidth]{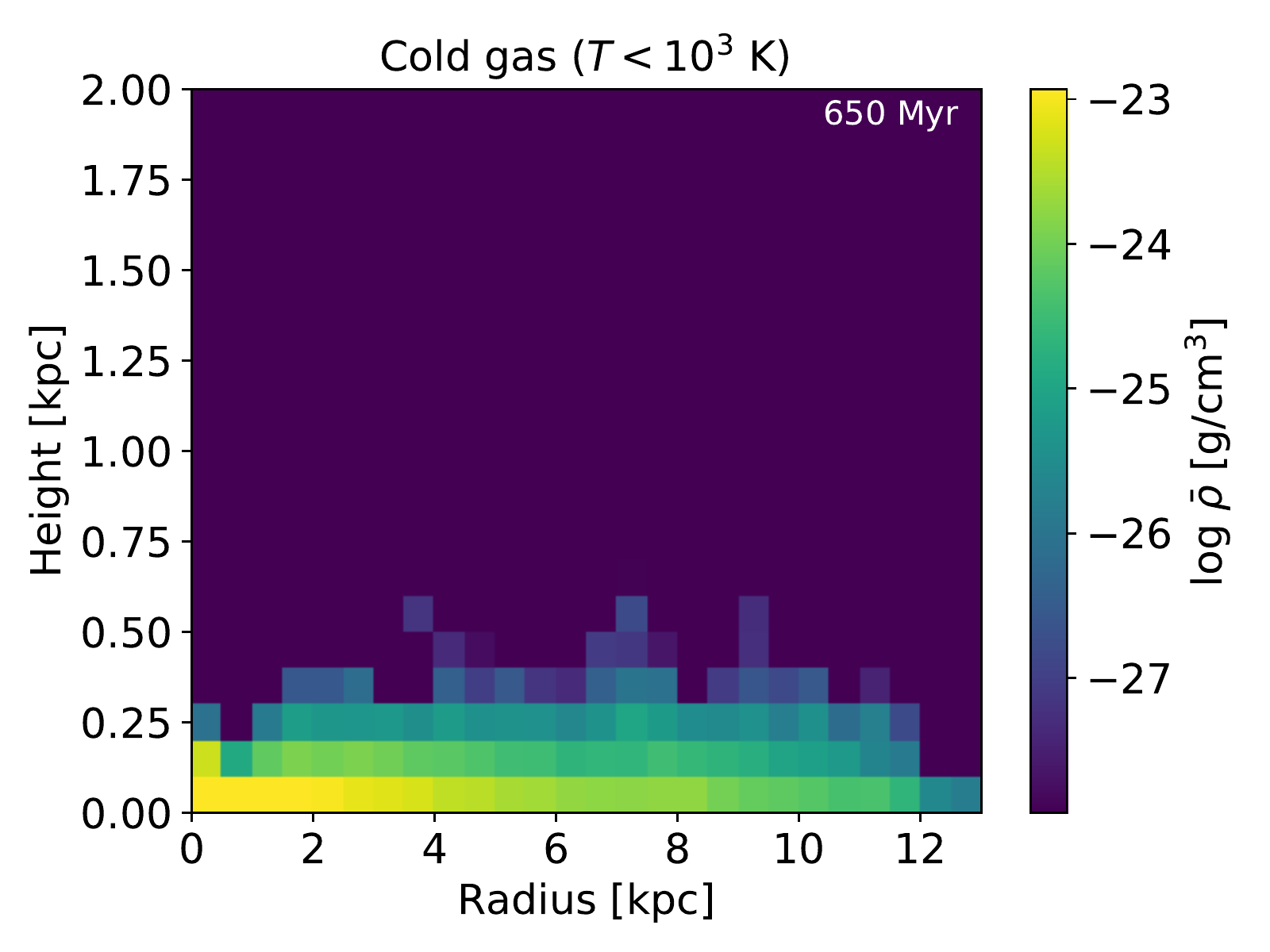}
	\includegraphics[width=0.9\figwidth]{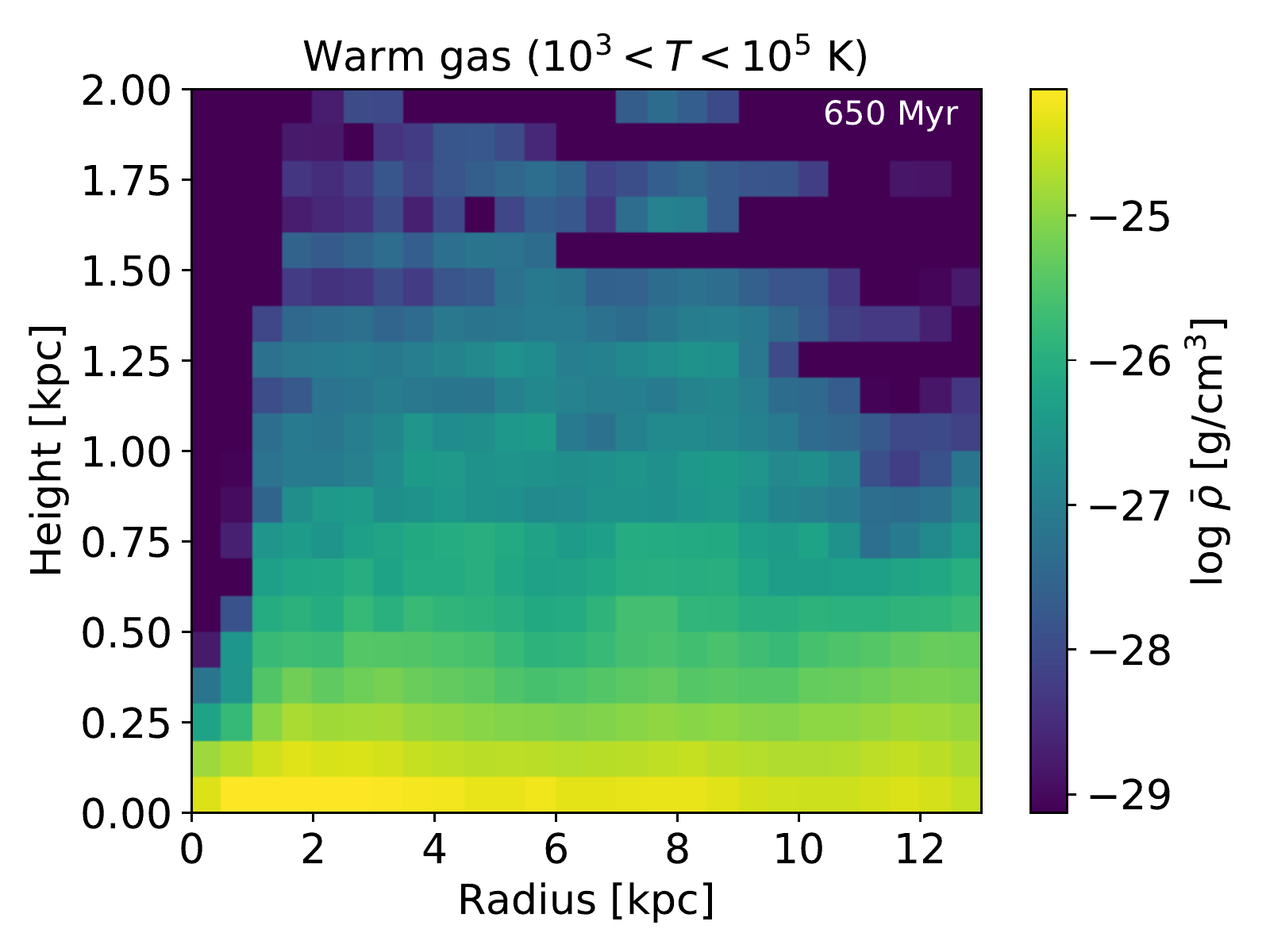}
	\includegraphics[width=0.9\figwidth]{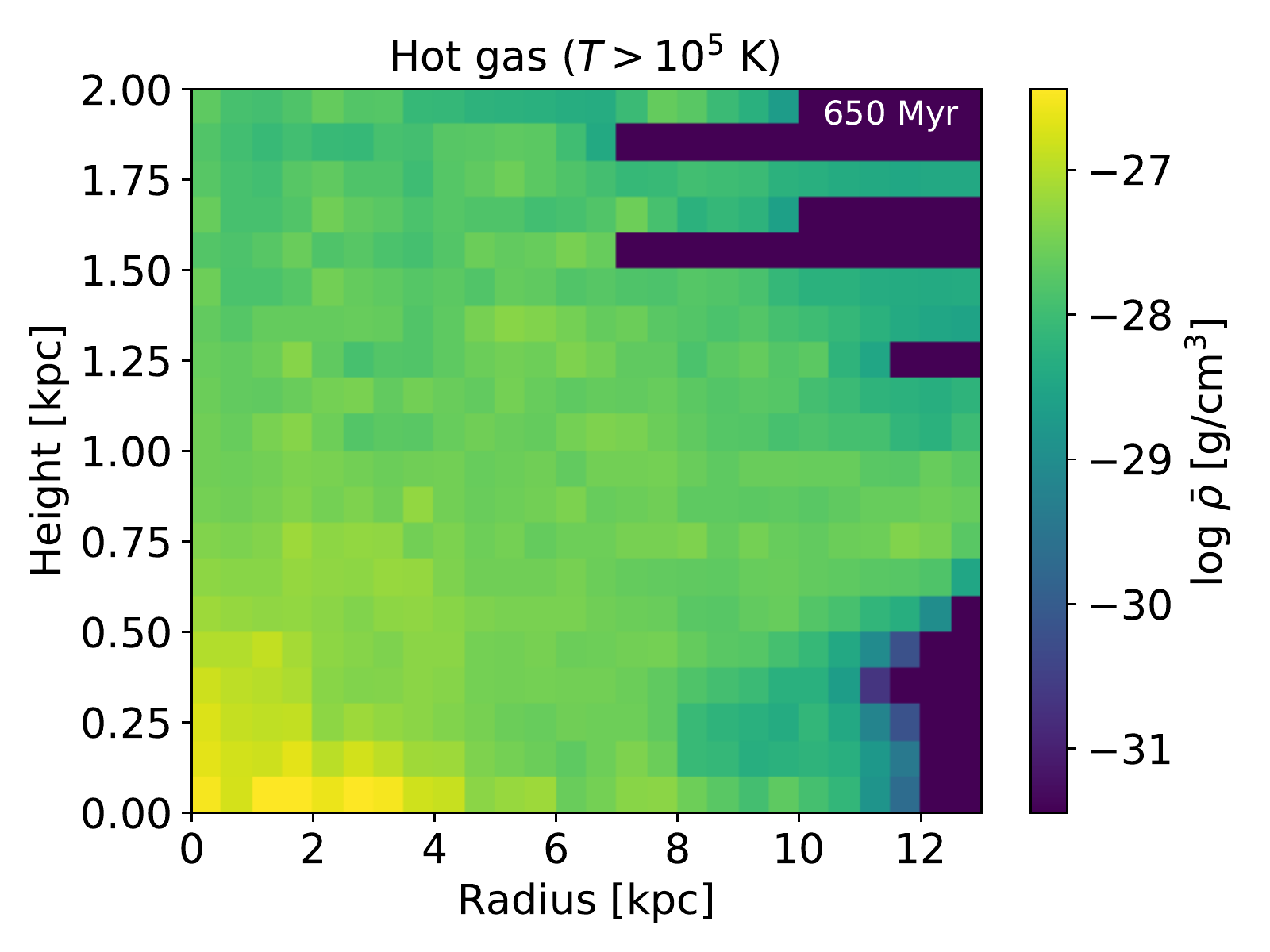}
    \caption{Radial and vertical distributions of $^{26}\mathrm{Al}$, cold, warm and hot-phase gas at $t = $ 650 Myr. Each bin shows the volume-weighted average density of material at a given radius and height, averaged in the azimuthal direction.}
    \label{fig: radius_vs_height}
\end{figure*}

Fig.~\ref{fig: radius_vs_height} shows azimuthally-averaged radial and vertical distributions of $^{26}\mathrm{Al}$ and of our three ISM temperature phases. The averages are over both sides of the disc, i.e., we have taken the mean of the points at $z>0$ and $z<0$. In the plot, each bin shows the volume-weighted average density.
We see that the distribution of $^{26}\mathrm{Al}$ is similar to that of the cold gas; most of the material is concentrated in the mid-plane of the galactic disc within $\left|\right.z\left.\right| < 250$ pc. Interestingly, $^{26}\mathrm{Al}$ shows a small portion distributed to higher altitudes that is not seen in the cold gas. On the other hand, the warm and hot gas are not confined within the mid-plane, and they are distributed to higher altitudes of $\left|\right.z\left.\right| > 1$ kpc. 

\begin{figure}
    \centering
	\includegraphics[width=\figwidth]{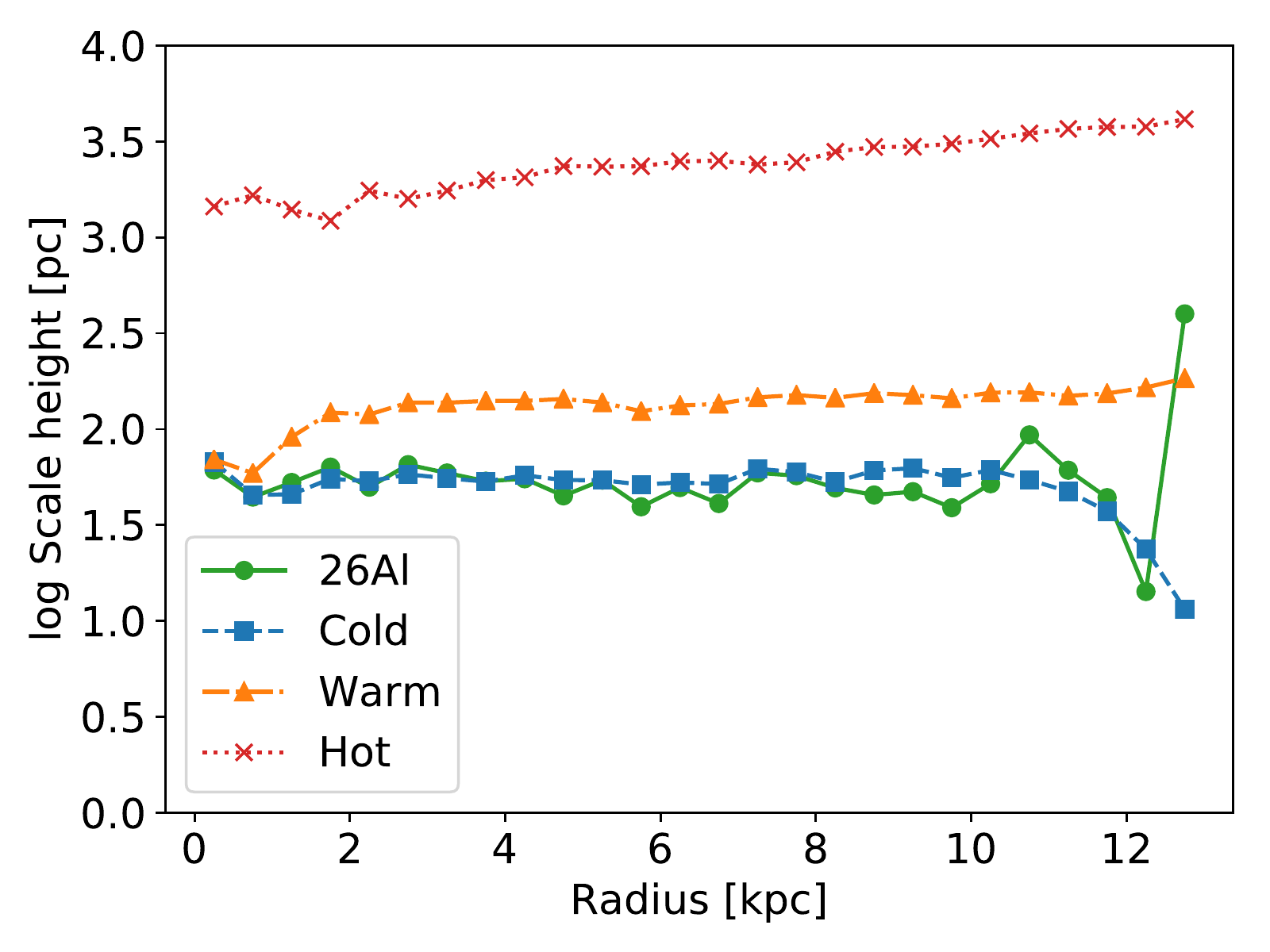}
    \caption{Scale heights of $^{26}\mathrm{Al}$, cold, warm and hot-phase gas as a function of galactocentric radius.}
    \label{fig: radius_vs_scale_height}
\end{figure}

We can make this discussion more quantitative by examining the scale height of each component, as shown in Fig.~\ref{fig: radius_vs_scale_height}. The scale height is defined as,
\begin{equation}
    H(R)  = \frac{\int_0^\infty \rho(R, z) z \mathrm{d}z}{\int_0^\infty \rho(R, z) \mathrm{d}z},
\end{equation}
where $R$ and $z$ are the cylindrical radial coordinate and the vertical coordinate, respectively, and all quantities are averaged over the azimuthal direction. The plot clearly shows small scale heights for $^{26}\mathrm{Al}$ and the cold gas, and large scale heights for the warm and hot gas.
The scale height of the cold gas is $\sim$ 50 pc, which is consistent with the scale height measured in Galactic CO surveys \citep{DameHartmannThaddeus2001}. On the other hand, the scale height of $^{26}\mathrm{Al}$ is one order of magnitude smaller than the estimates from $\gamma$-ray observations \citep{WangEtAl2020}, even though this simulation shows prominent spiral arms and large star-forming regions along them. \citet{Pleintinger2019} compared our previous simulation, \citet{FujimotoKrumholzTachibana2018}, which did not include stellar spiral arms, with $\gamma$-ray observations, and found a similar result: the scale height in the simulated $^{26}\text{Al}$ disc is much smaller than that inferred from observations. They speculated that one possible explanation for this discrepancy could be lack of sufficiently strong spiral arms that could foster massive molecular clouds and thus large star-forming regions that could spread $^{26}\mathrm{Al}$ to higher altitudes. However, the small scale height of $^{26}\mathrm{Al}$ shown in Fig.~\ref{fig: radius_vs_scale_height} indicates that the scale height does not change substantially if we do self-consistently include spiral arms whose properties are consistent with our current understanding of the Milky Way's arms. This appears to invalidate the hypothesis expressed in \citet{Pleintinger2019}.

\begin{figure*}
    \centering
	\includegraphics[width=0.9\figwidth]{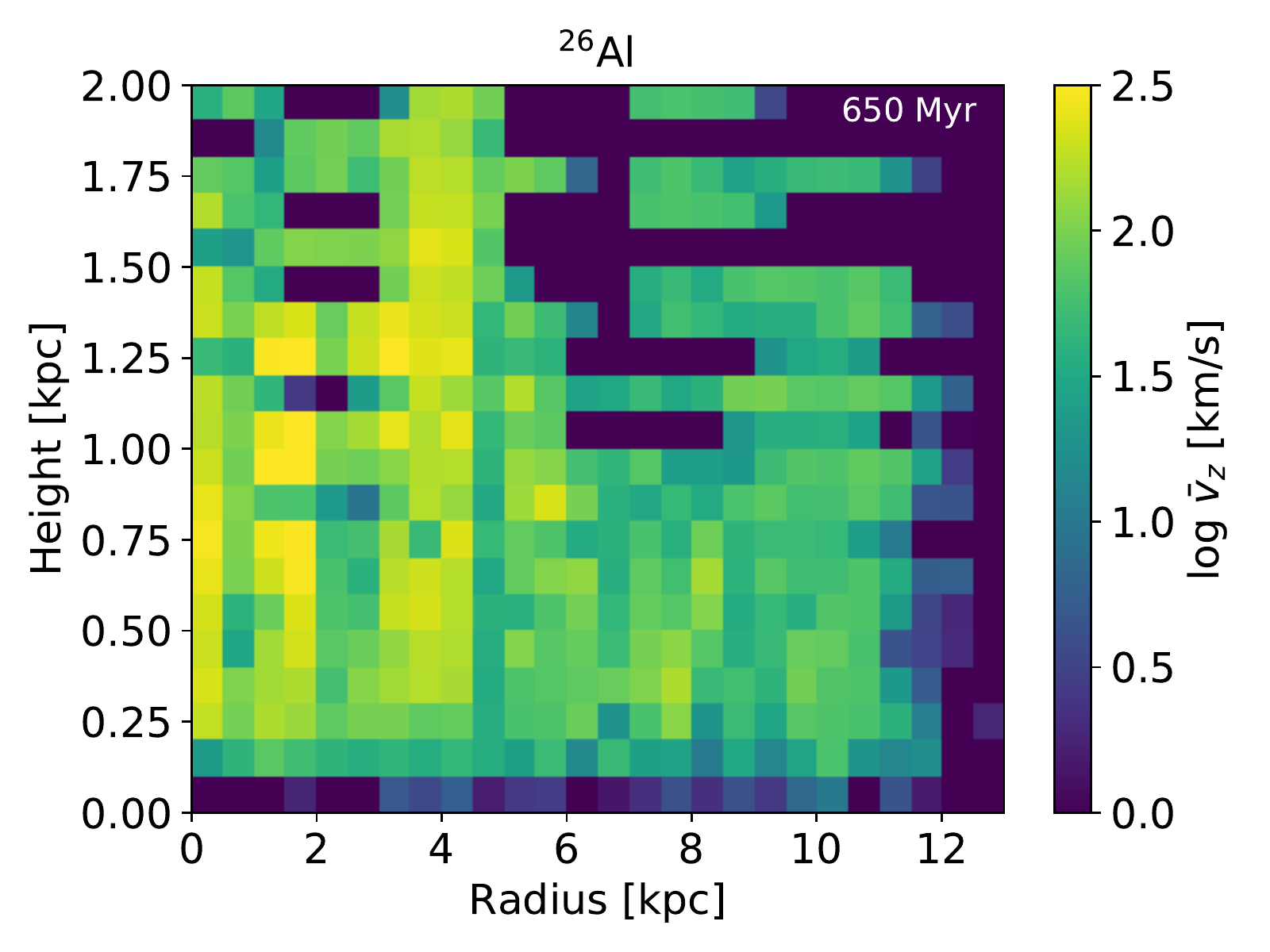}
	\includegraphics[width=0.9\figwidth]{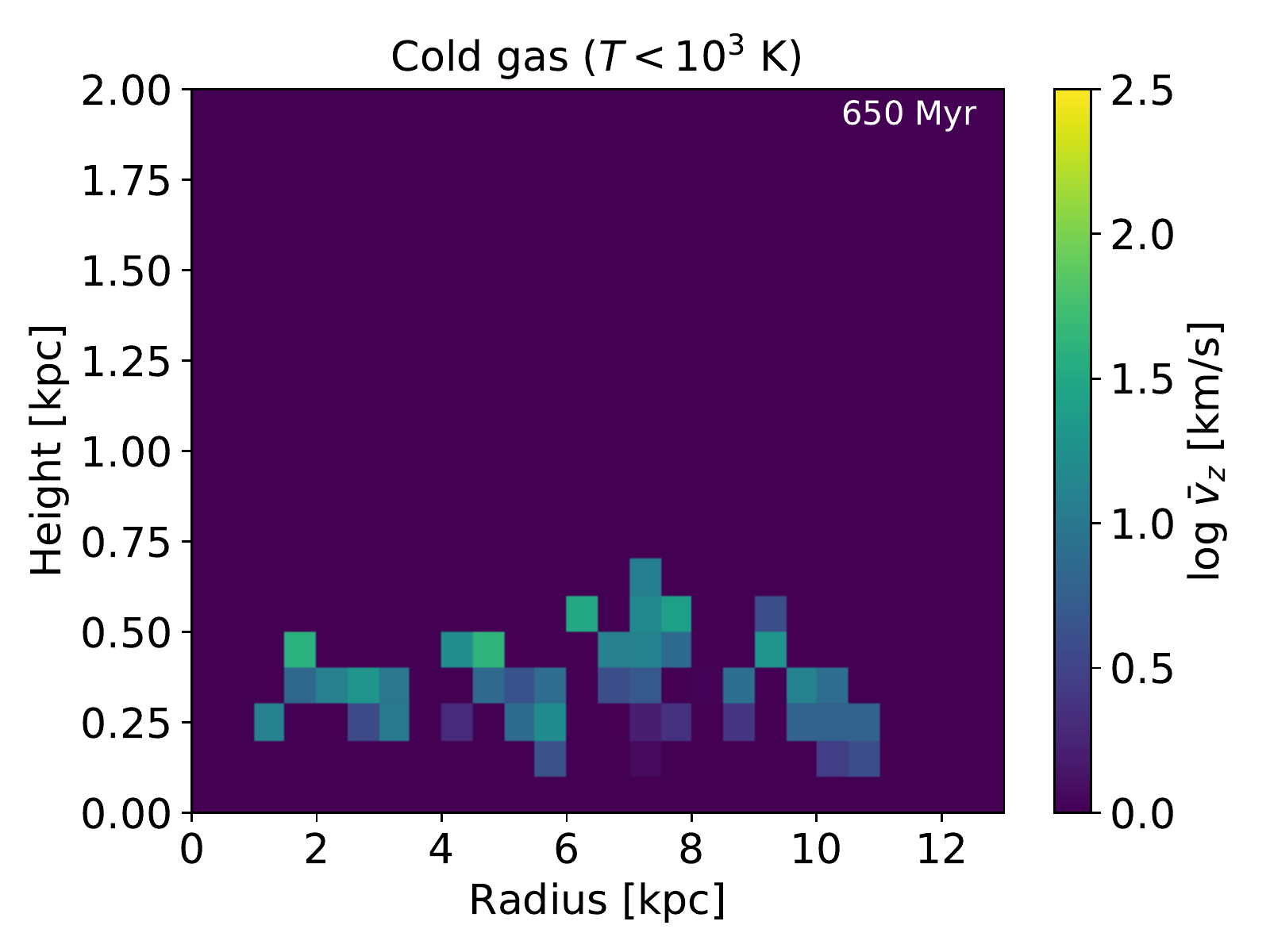}
	\includegraphics[width=0.9\figwidth]{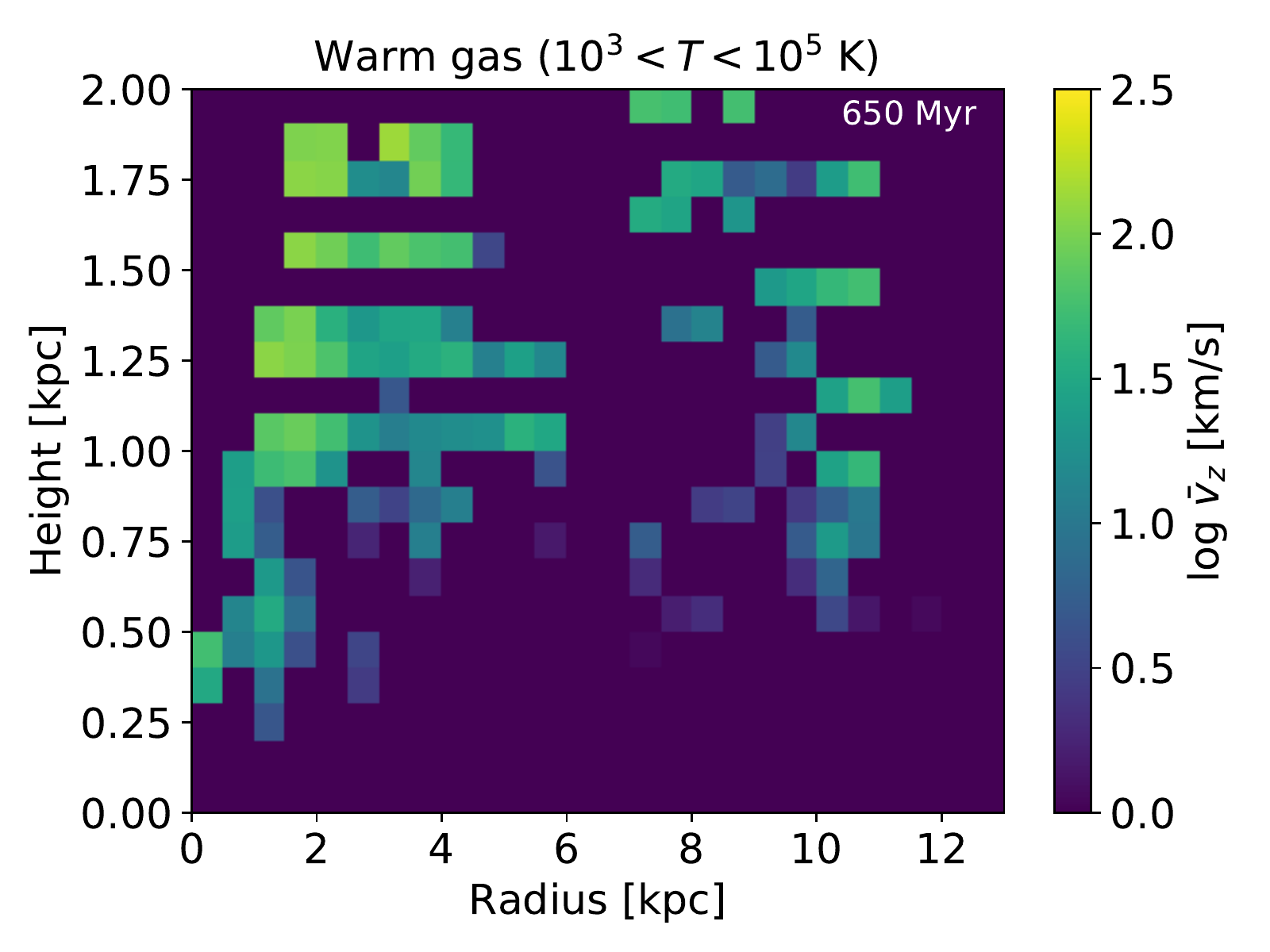}
	\includegraphics[width=0.9\figwidth]{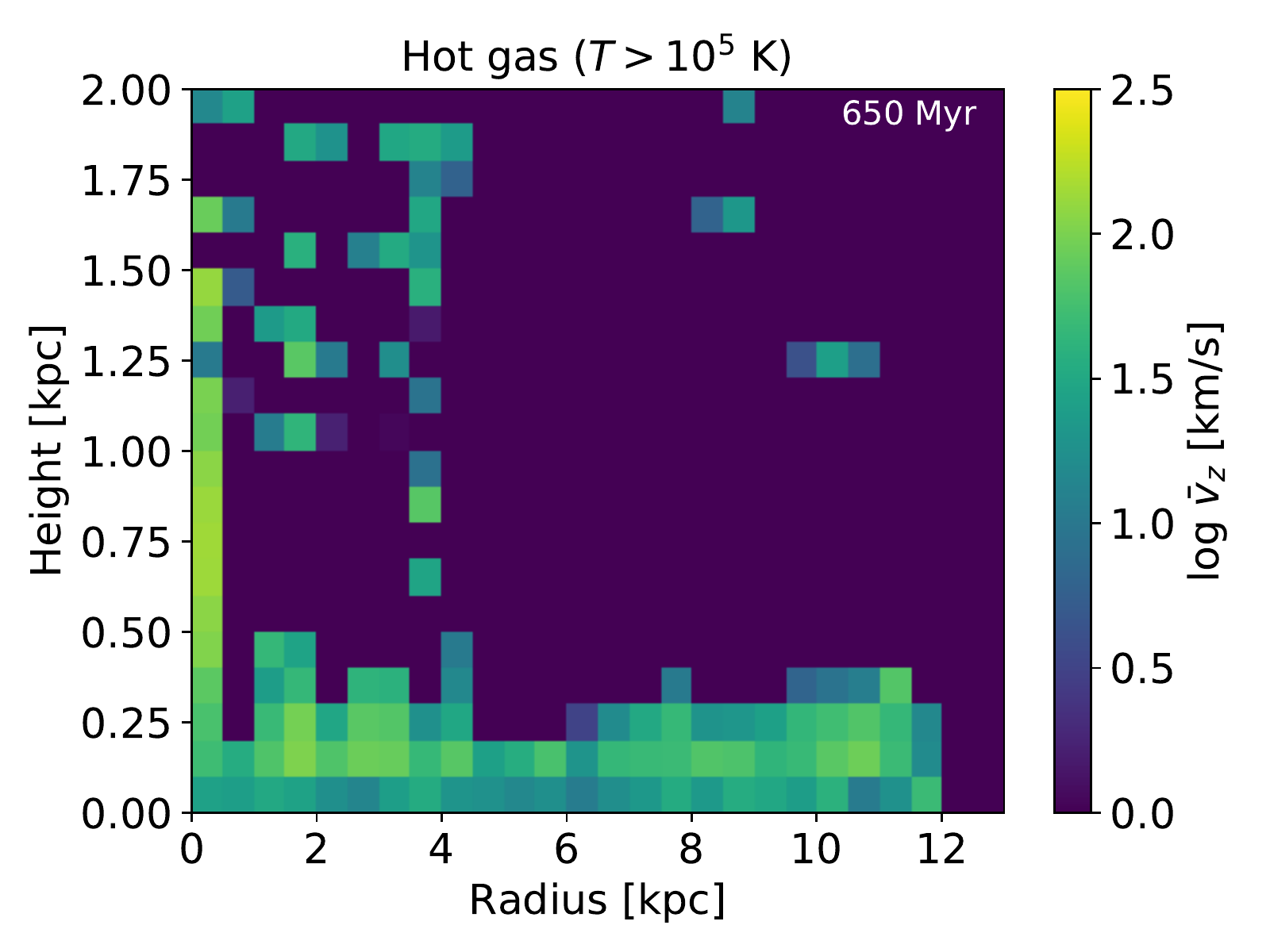}
    \caption{Outflow velocities from the mid-plane of the galactic disc with respect to height versus radius. Each bin shows the mass-weighted average for $^{26}\mathrm{Al}$, cold, warm and hot-phase gas.}
    \label{fig: radius_vs_height_with_vz}
\end{figure*}

While the scale height of the $^{26}\mathrm{Al}$ is clearly not hundreds of pc, it is noteworthy, and an important clue to which we shall make reference later, that a part of the $^{26}\mathrm{Al}$ distribution shown in Fig.~\ref{fig: radius_vs_height} is distributed at high altitudes; there is no analogous feature in the cold gas. To understand the origin of this gas, in Fig.~\ref{fig: radius_vs_height_with_vz} we show the mass-weighted mean outflow velocity of $^{26}\text{Al}$ and of our three ISM temperature phases, averaged in bins of height and radius as in Fig.~\ref{fig: radius_vs_height}. Here we define the outflow velocity relative to the disc as $\max(\hat{z}\cdot \mathbf{v},0)$, i.e., the outflow velocity is the component of the velocity vector that points away from the galactic plane. The figure clearly shows that a large amount of $^{26}\mathrm{Al}$ has large outflow velocities, and that this is not the case in the cold gas; it shows that, although $^{26}\mathrm{Al}$ has a small scale height similar to that of the cold gas, this is because of its short half-life, not its kinematics, which are more similar to those of the warm or hot phases. As shown in Fig.~\ref{fig: radius_vs_height_with_vz}, at heights above $\sim 100$ pc, only $^{26}\mathrm{Al}$ with velocities larger than $\sim 100$ $\mathrm{km\ s^{-1}}$, is present, which makes sense: material travelling at $<100$ km s$^{-1}$ will travel $<100$ pc over the $\approx 1$ Myr half-life of $^{26}\text{Al}$, so only material launched at larger velocities will have time to reach large heights. Indeed, this effect might also explain the difference between the $^{26}\text{Al}$ velocity distribution and that of the warm and hot phases. Because these phases take much longer than 1 Myr to ``decay'' (via cooling), their velocity fields are dominated by contributions from the accumulated gas ejected over many star formation events, of which has had time to settle into an equilibrium state. Due to its shorter lifetime, the $^{26}\mathrm{Al}$ more closely follows the kinematics of newly-ejected gas from young massive stars.

\subsection{Rotation of $^{26}\mathrm{Al}$ in the disc}
\label{sec: Rotation of 26Al in the disc}

\begin{figure*}
    \centering
	\includegraphics[width=0.9\figwidth]{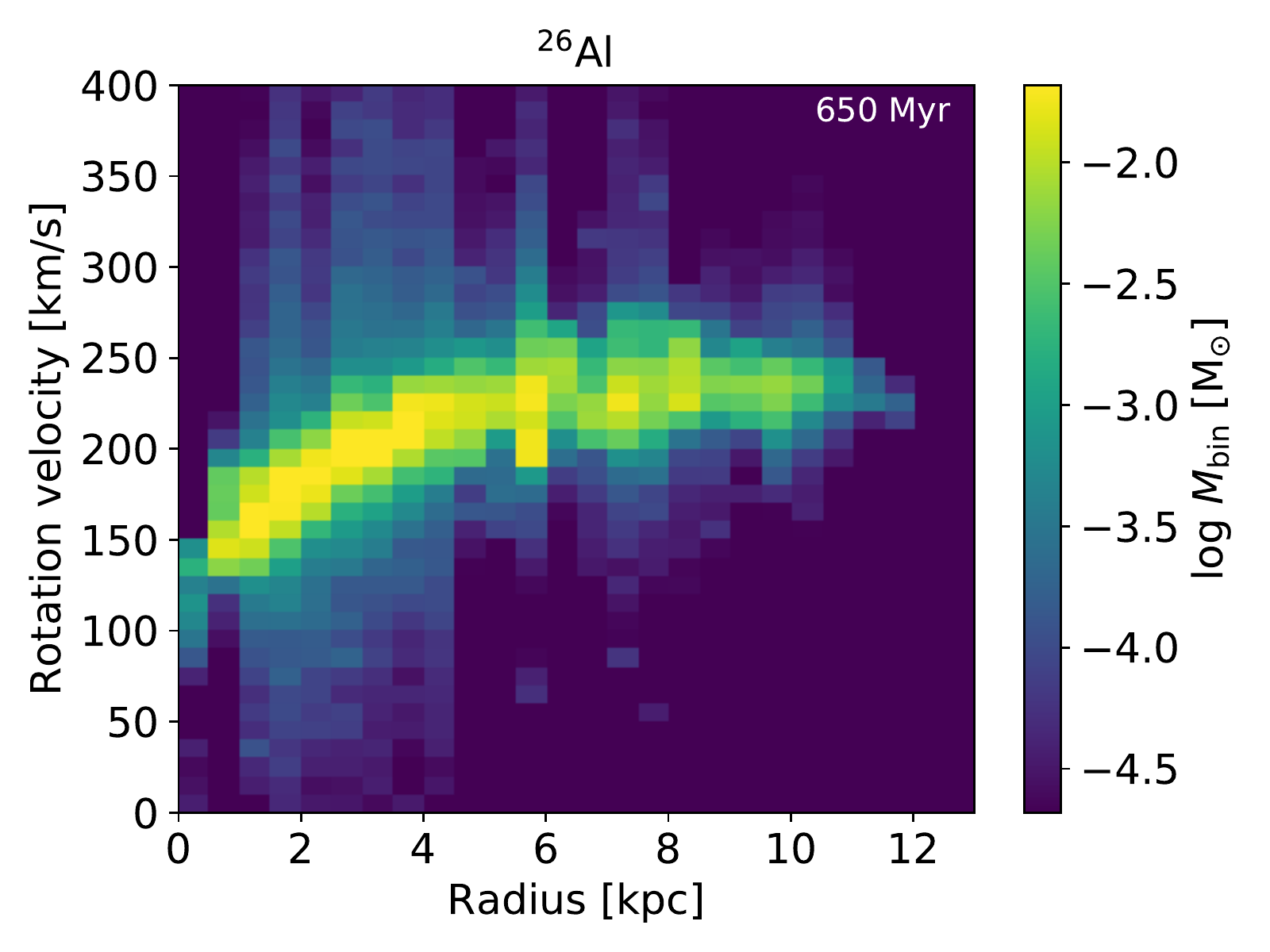}
	\includegraphics[width=0.9\figwidth]{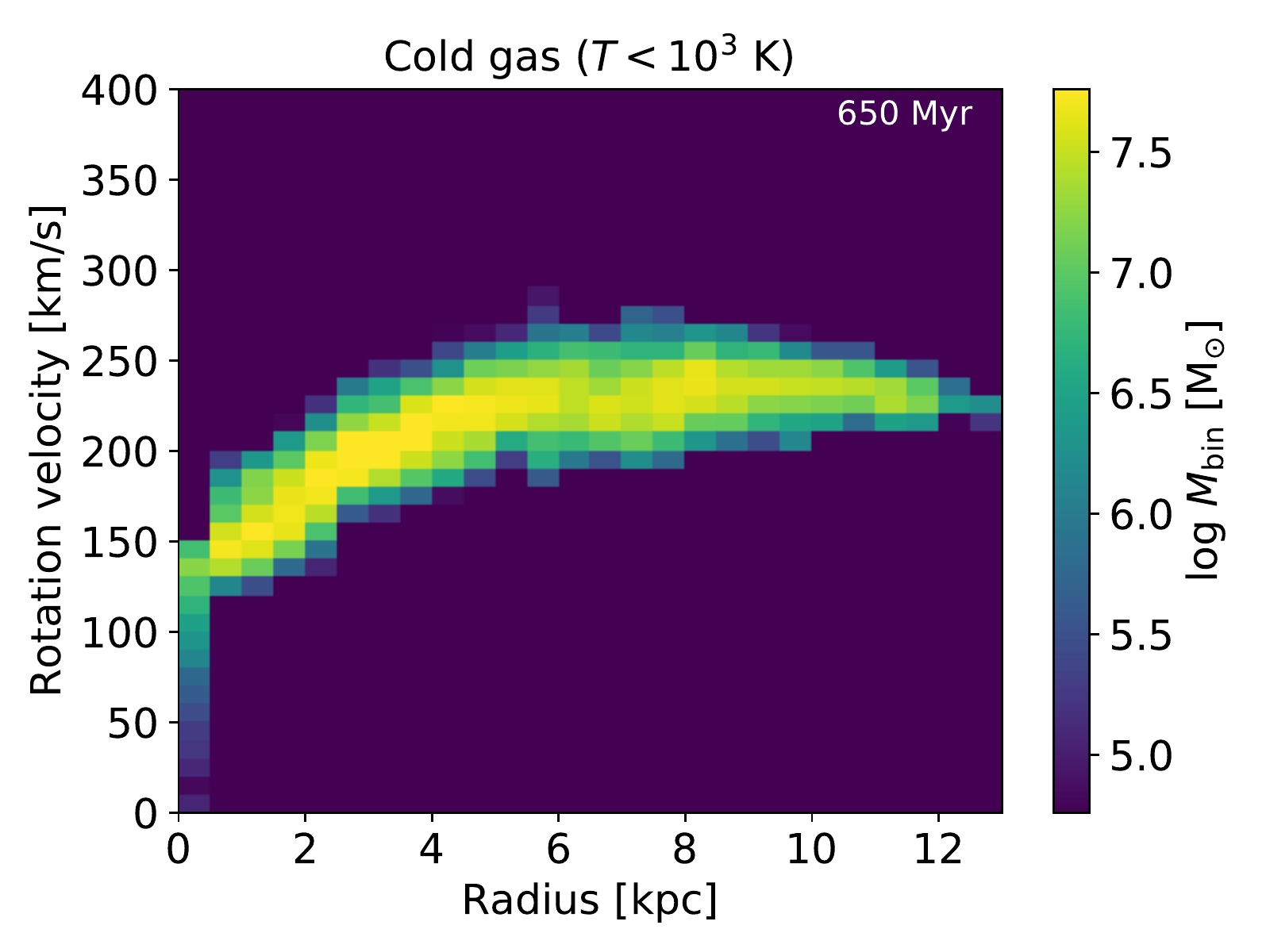}
	\includegraphics[width=0.9\figwidth]{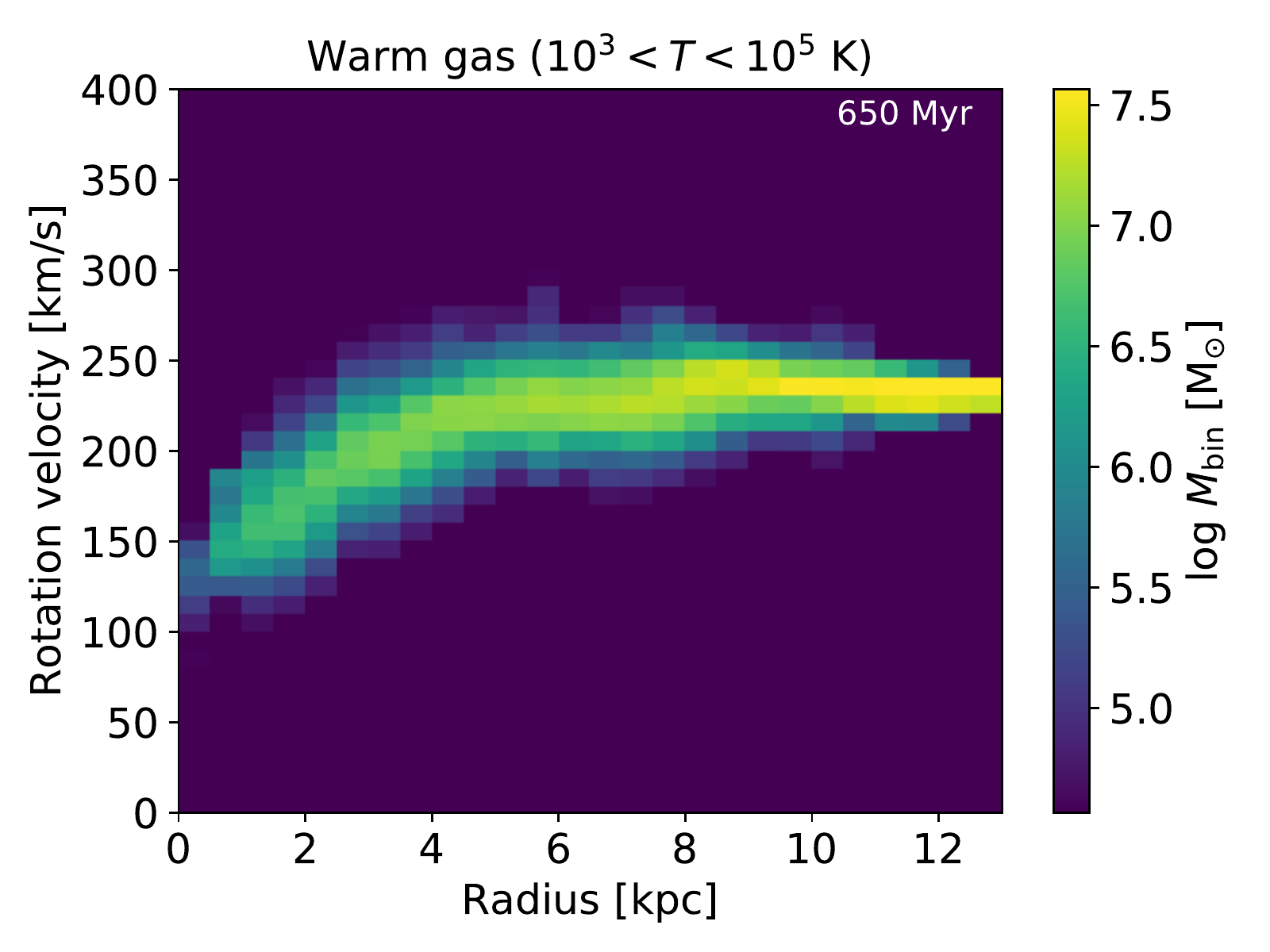}
	\includegraphics[width=0.9\figwidth]{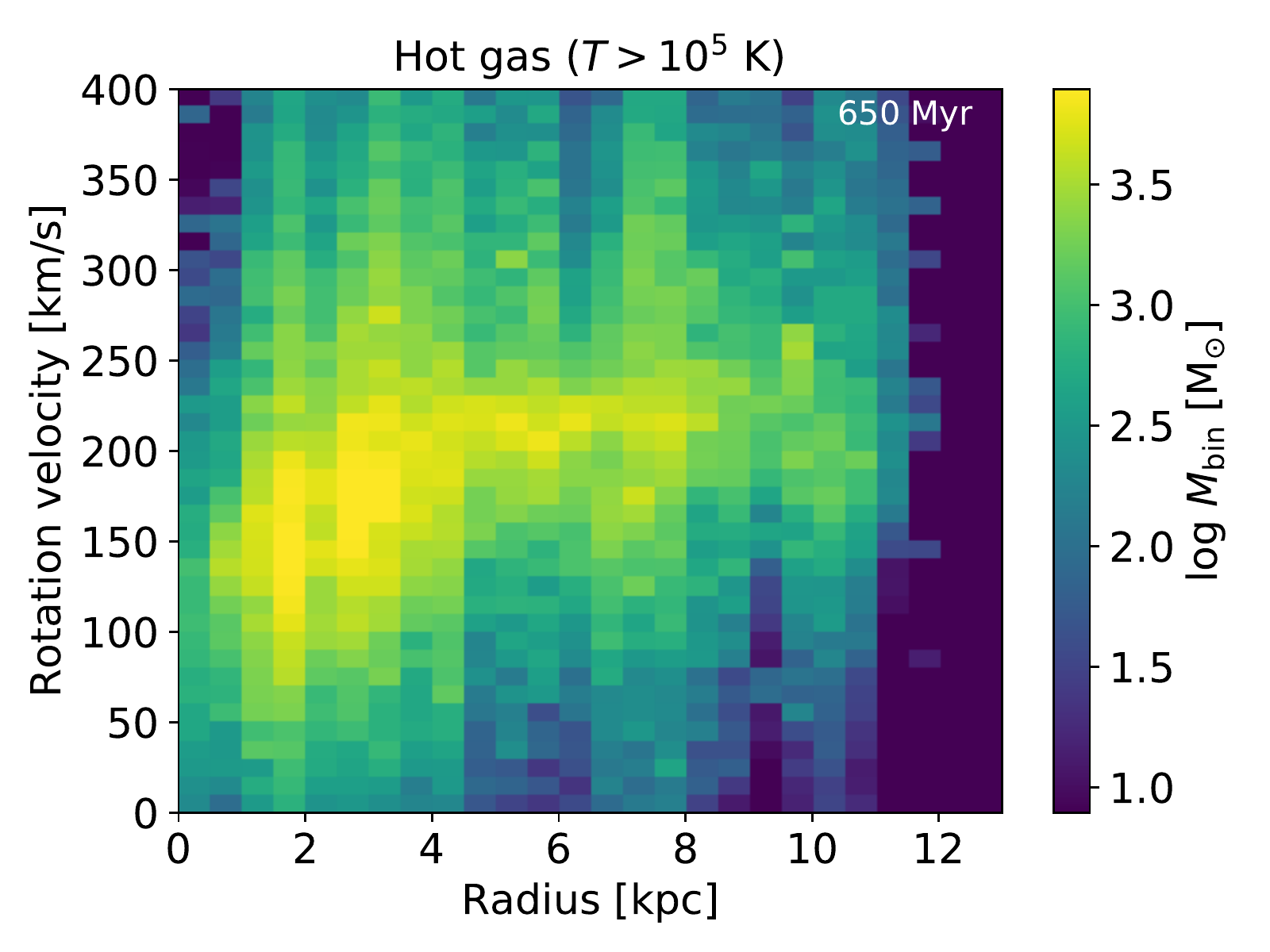}
    \caption{Mass distributions with respect to rotation velocity versus radius. Each bin shows the sum of mass for $^{26}\mathrm{Al}$, cold, warm and hot-phase gas integrated over all cells in the galactic disc within $-250\ \mathrm{pc} < z < 250\ \mathrm{pc}$.}
    \label{fig: radius_vs_circular_velocity}
\end{figure*}

Our final step in analysing the simulations is to examine the rotation of the gas. Fig.~\ref{fig: radius_vs_circular_velocity} shows distribution of mass in bins of radius and rotation velocity for each ISM component, for gas within 250 pc of the mid-plane. Although there are some excesses, the mean rotation velocities at each radius for $^{26}\mathrm{Al}$ are similar to those for the cold and warm gas. The hot gas shows a very large range of variation, but the mean is nonetheless consistent with the other phases. We see no evidence for systematically faster rotation of the $^{26}\text{Al}$ compared to the bulk of the gas. The $\gamma$-ray observations, on the other hand, have shown a different picture; the rotation velocities of $^{26}\mathrm{Al}$ have a systematic excess up to 200 $\mathrm{km\ s^{-1}}$ compared to those of the cold ISM \citep{KretschmerEtAl2013}. We discuss possible reasons in the following section.

\section{Discussion}
\label{sec: Discussions}

Having analysed our simulation results, we now consider their implications for our two possible scenarios for why the observed $^{26}\text{Al}$ position-velocity distribution differs strongly from that of the massive stars or the gas from which they form.

\subsection{Possibility 1: galactic spiral structure}
\label{sec: Possibility 1: galactic spiral structure}

As discussed in the introduction, \citet{KretschmerEtAl2013} and \citet{KrauseEtAl2015} propose that the observation that $^{26}\mathrm{Al}$ rotates faster than the gas in the Milky Way disc can be explained if massive star clusters form at the leading edge of the gaseous spiral arms. These clusters would then form superbubbles filled with hot-phase ISM that would expand to the forward direction against galactic rotation, while also breaking out to large scale height, due to a hydrodynamical interaction with the locally anisotropic ISM of the gaseous spiral arm \citep[see Figure 4 in][]{KrauseEtAl2015}. \citet{Rodgers-LeeEtAl2019} perform hydrodynamic simulations of this scenario, including a subgrid superbubble model and an external spiral arm potential that rotates in the disc with a fixed pattern speed. Although there are still some quantitative discrepancies with observations, they succeed in reproducing some qualitative features: a large $^{26}\mathrm{Al}$ scale height of 5 kpc and a velocity excess up to 200 $\mathrm{km\ s^{-1}}$. 

This scenario depends strongly on the subgrid superbubble recipe, which is substantially uncertain due to the inability of current simulations to resolve the conductive interface between hot and cool ISM phases -- e.g., see \citet{GentryEtAl2017, GentryEtAl2019} versus \citet{Kim17a} and \citet{El-Badry19a}. While our simulations also lack the resolution to address this uncertainty, they do allow us to address another requirement of this scenario: that clusters form along the leading edges of spiral arms, so that their SN bubbles blow out preferentially in the forward direction. Whether this requirement is met in the Milky Way depends on the formation mechanism for the Milky Way's spiral arms, a subject that has been debated for long time \citep[see][for a review; see also \citealt{SellwoodCarlberg2019}]{DobbsBaba2014}. The spiral structure proposal for explaining $^{26}\text{Al}$ kinematics implicitly (and explicitly in the case of \citeauthor{Rodgers-LeeEtAl2019}'s simulations) assumes that the arms are density waves \citep{LinShu1964, BertinLin1996}: long-lived quasi-stationary structures that propagate azimuthally through the galactic disc at a constant pattern speed. Arms of this type lead to a clear offset between gas and stars across the arm caused by a shock on the forward side of the arm \citep{BabaMorokumaMatsuiEgusa2015}, as required by the spiral structure proposal. Such offsets are in fact seen in some grand design spirals, such as M51 and M81, that are undergoing clear tidal interactions \citep[e.g.,][]{PettittTaskerWadsley2016}.

However, offsets between gas, stars, and star formation tracers are generally \textit{not} found in galaxies that lack grand design structures \citep[e.g.,][]{EgusaEtAl2009, FoyleEtAl2011, FerrerasEtAl2012}. Simulations suggest that, in these galaxies, the spiral features are likely short-lived and non-stationary, and are characterised by multiple arms that continuously appear, merge, and shear away \citep[e.g.][]{WadaBabaSaitoh2011, FujiiEtAl2011, GrandKawataCropper2012a, GrandKawataCropper2012b, DOnghiaVogelsbergerHernquist2013, BabaSaitohWada2013, SellwoodCarlberg2014}. Arms of this type are material arms rather than standing waves, so there is no front-back asymmetry in the location of star formation relative to the arm. While our location within the Milky Way means that we cannot perform precisely the same observations for it as for the extragalactic systems, there is strong evidence that the Milky Way falls into the category of galaxies with transient arms rather than semi-permanent density waves. This evidence includes numerical $N$-body simulations showing that transient arms better reproduce Milky Way's stellar velocity fields observed with the \textit{APOGEE} and \textit{Gaia} surveys \citep[][]{KawataEtAl2014, GrandEtAl2015, BabaEtAl2018, HuntEtAl2018, SellwoodEtAl2019}, hydrodynamic simulations showing that transient arms best match the observed CO distribution \citep{PettittEtAl2015}, and age distributions of stars within 3 kpc of the Sun from \textit{Gaia} DR2 data that suggest arms lifetimes of no more than a few hundred Myr \citep{KounkelCovey2019, KounkelCoveyStassun2020}.

Our simulation in this work assumes the transient arm scenario favoured by this evidence, in that we do not include either a fixed spiral potential or a companion capable of inducing a grand design pattern. It shows that, in this case, there is no systematic preference for star cluster formation to be on the leading edge of the arms, and thus no asymmetric blowout of SN bubbles in the forward direction. This suggests that the spiral structure scenario for $^{26}\text{Al}$ kinematics is likely incorrect. However, we do raise one caution: although both recent numerical and observational studies suggest that the Milky Way's arms are transient and material rather than quasi-static density waves, it is still possible that perturbations from companions and/or interactions with the Galactic bar (which also drive grand design features near the bar tips -- \citealt{GrandKawataCropper2012b, Baba2015}) enhance the Milky Way's spiral arms, and then that the properties of the spiral arms become density wave-like. 
Recently, it has been suggested that not all aspects of distribution in position and velocity of nearby stars from the \textit{Gaia} DR2 can be explained by the transient material arm model, and some data are compatible with the rigidly rotating spiral arm model \citep{SellwoodEtAl2019, PettittRaganSmith2020}. In addition, perturbations to the spiral structure of the Milky Way due to a recent passage of the Sagittarius dwarf galaxy through the Milky Way disc, and resulting spatial offsets of gas and young stars across the arm were identified \citep{LaporteEtAl2019, Bland-HawthornEtAl2019}.
In fact, the observed lifetime of the Local Arm is $\sim$ 300 Myr \citep{KounkelCovey2019, KounkelCoveyStassun2020}, which is slightly longer than that seen in our simulation ($\sim$ 100 Myr). It suggests that the Milky Way's arms are somewhat enhanced by companions and/or the galactic bar.
The effects of passing companions and/or the galactic bar on the $^{26}\mathrm{Al}$ distribution and kinematics should be investigated in future work.

\subsection{Possibility 2: foreground local structures}
\label{sec: Possibility 2: foreground local structures}

\begin{figure*}
    \centering
	\includegraphics[width=0.9\figwidth]{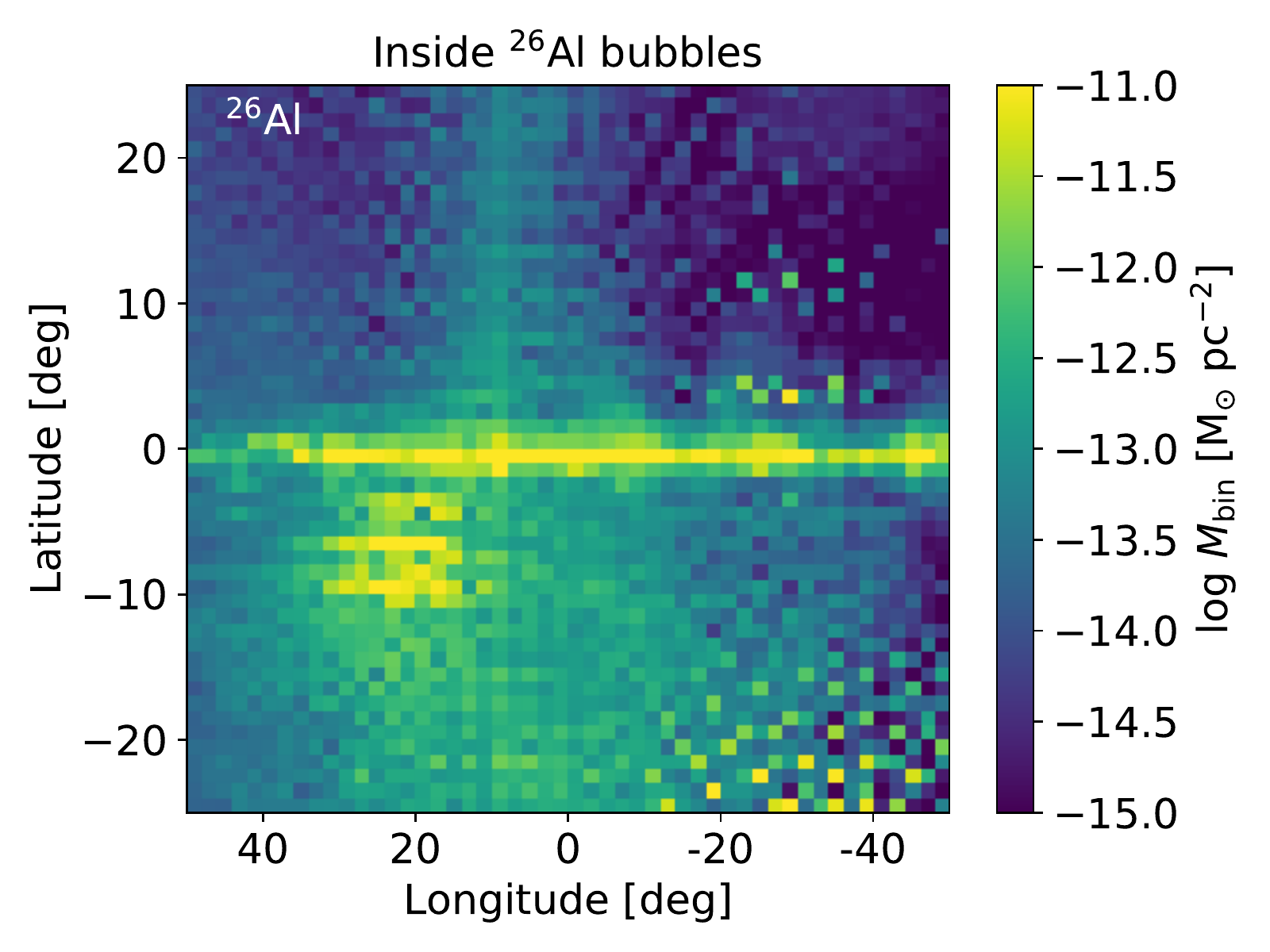}
	\includegraphics[width=0.9\figwidth]{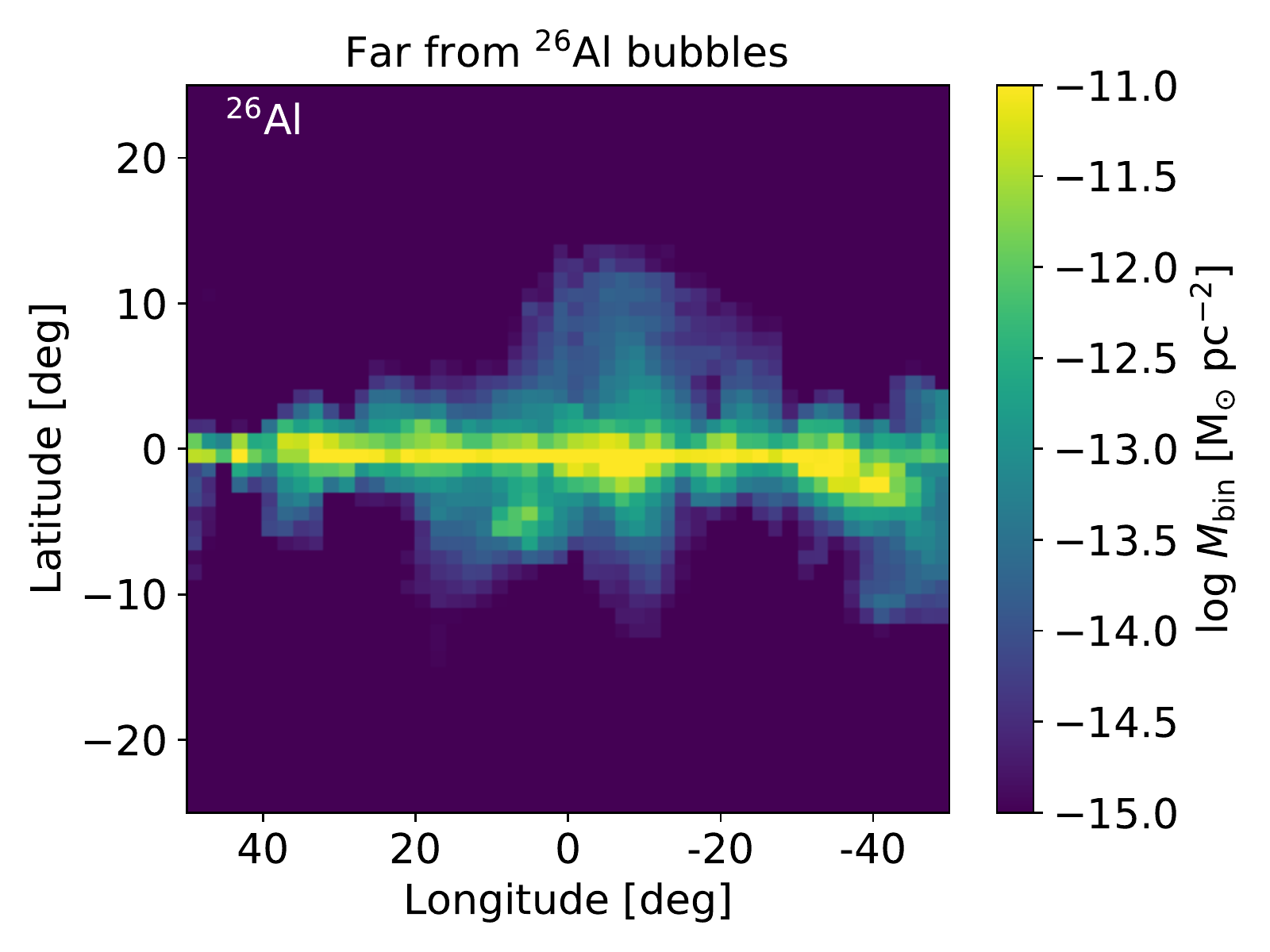}
	\includegraphics[width=0.9\figwidth]{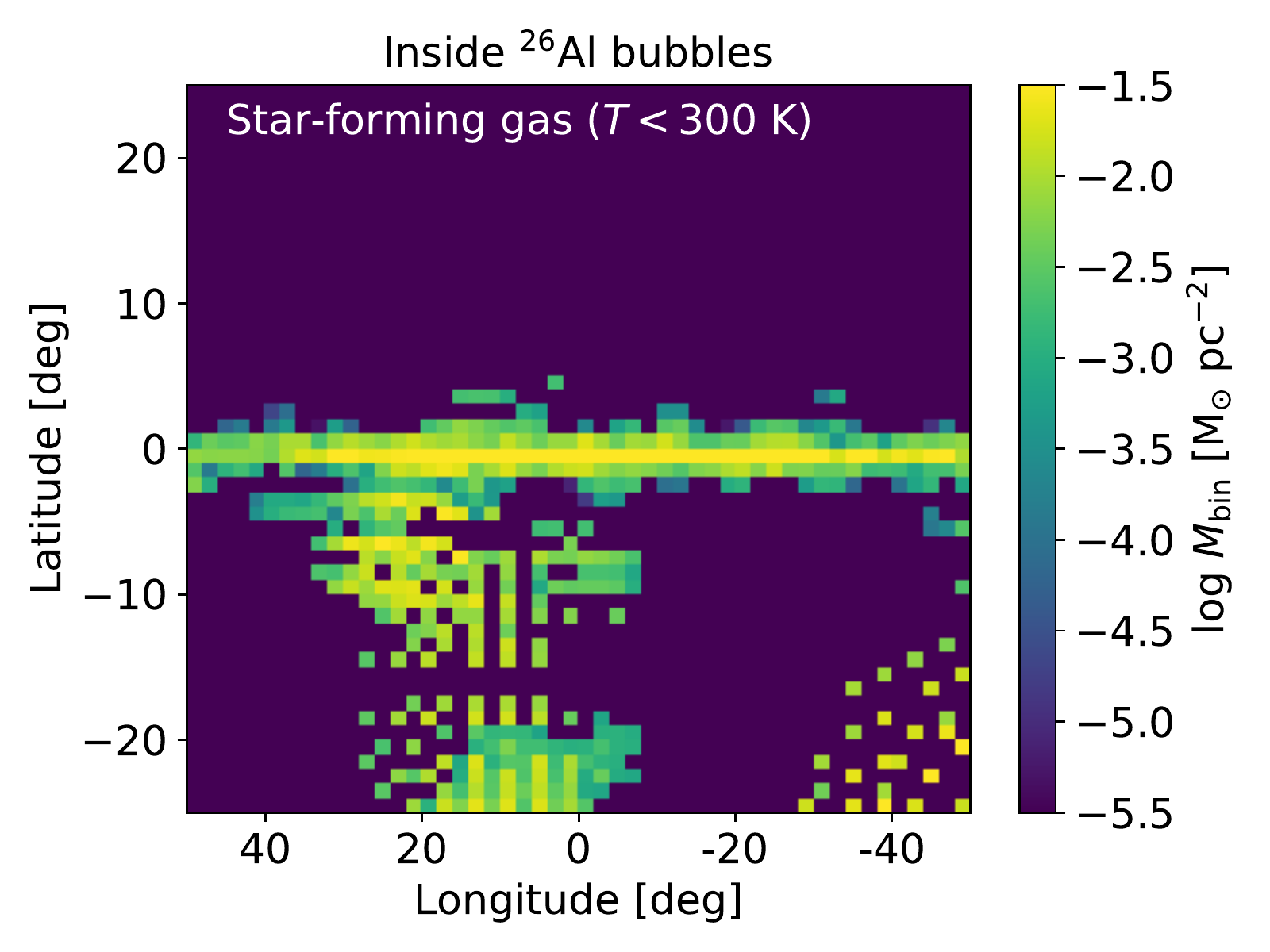}
	\includegraphics[width=0.9\figwidth]{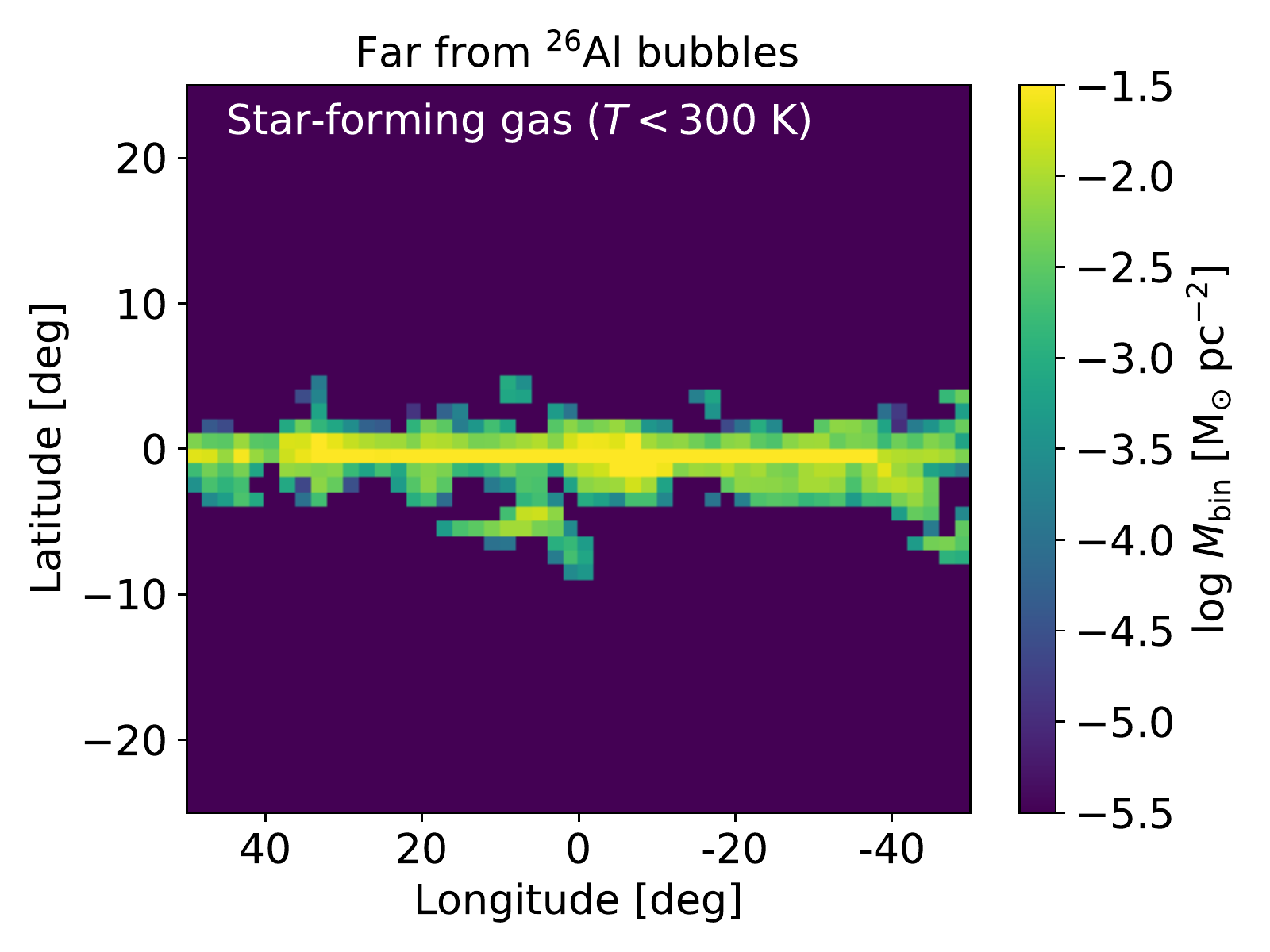}
	\includegraphics[width=0.9\figwidth]{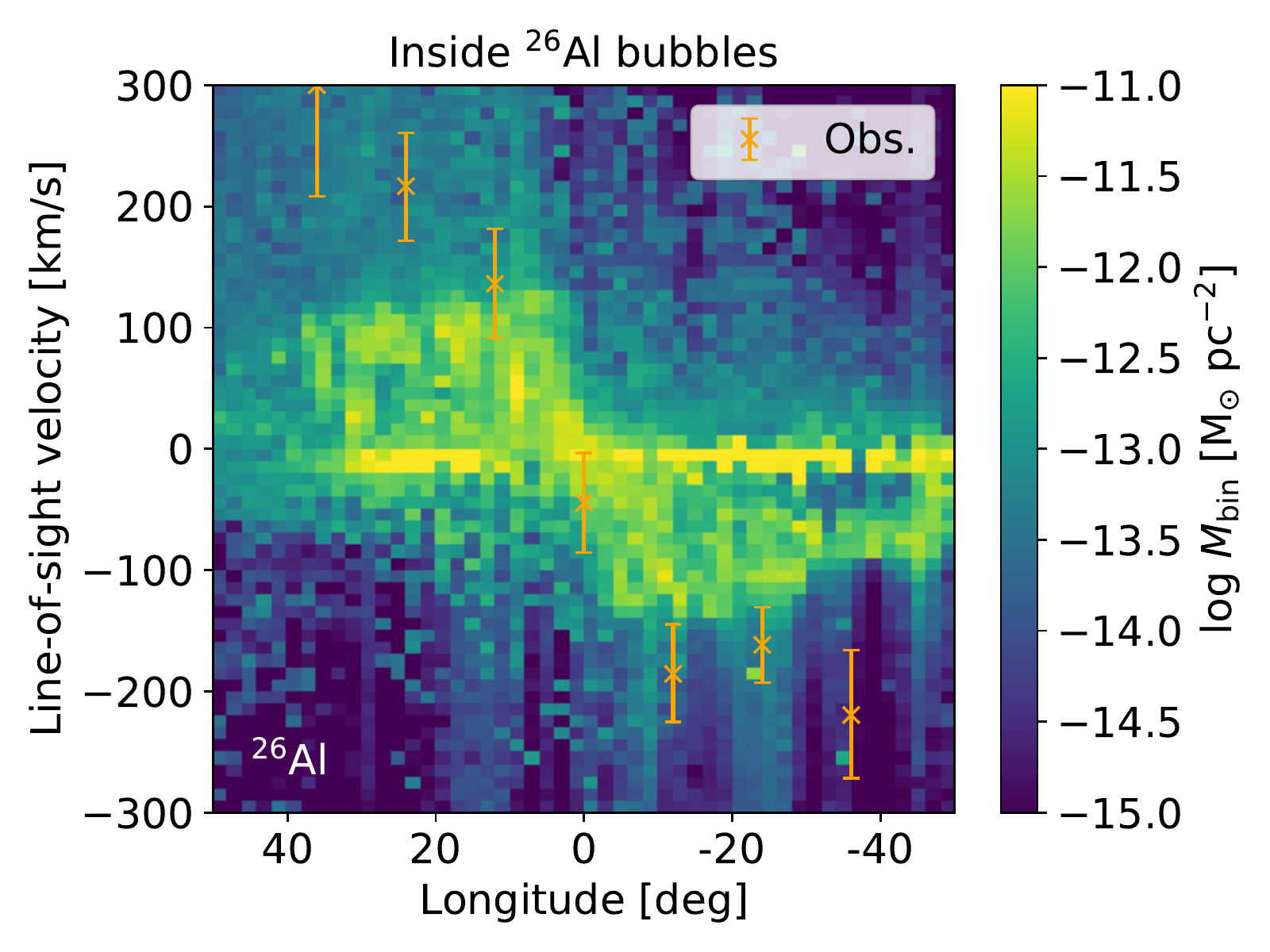}
	\includegraphics[width=0.9\figwidth]{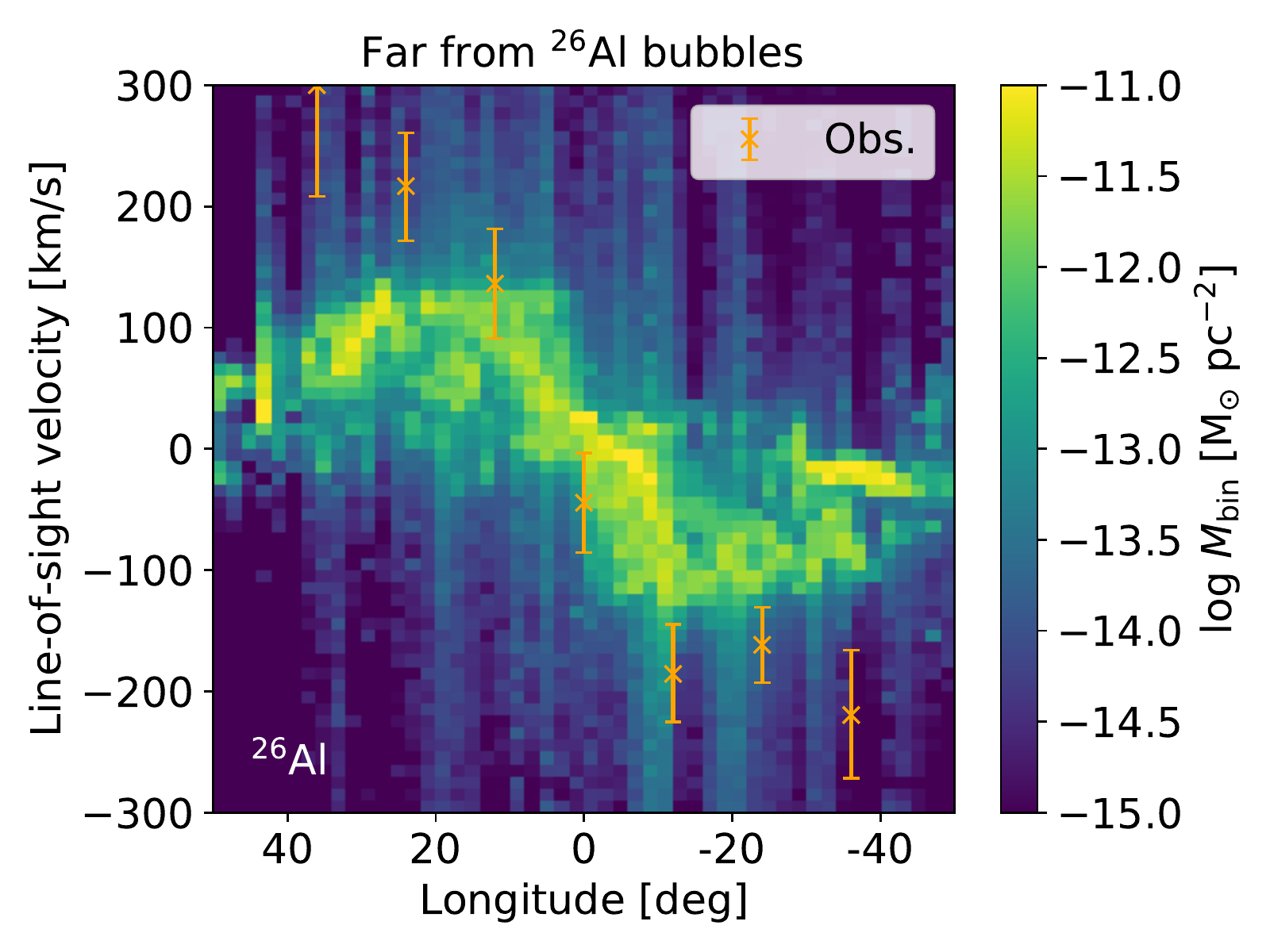}
	\includegraphics[width=0.9\figwidth]{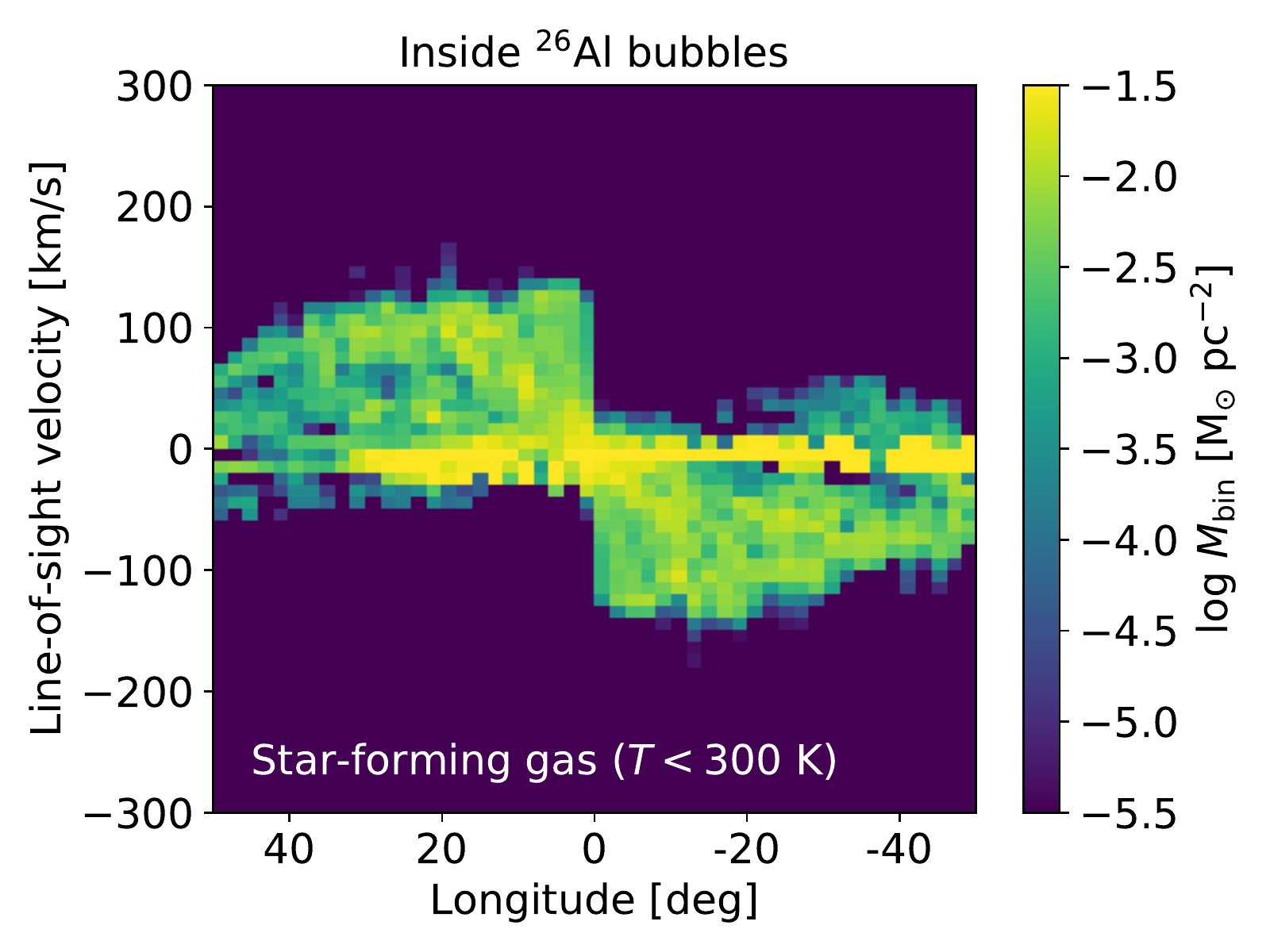}
	\includegraphics[width=0.9\figwidth]{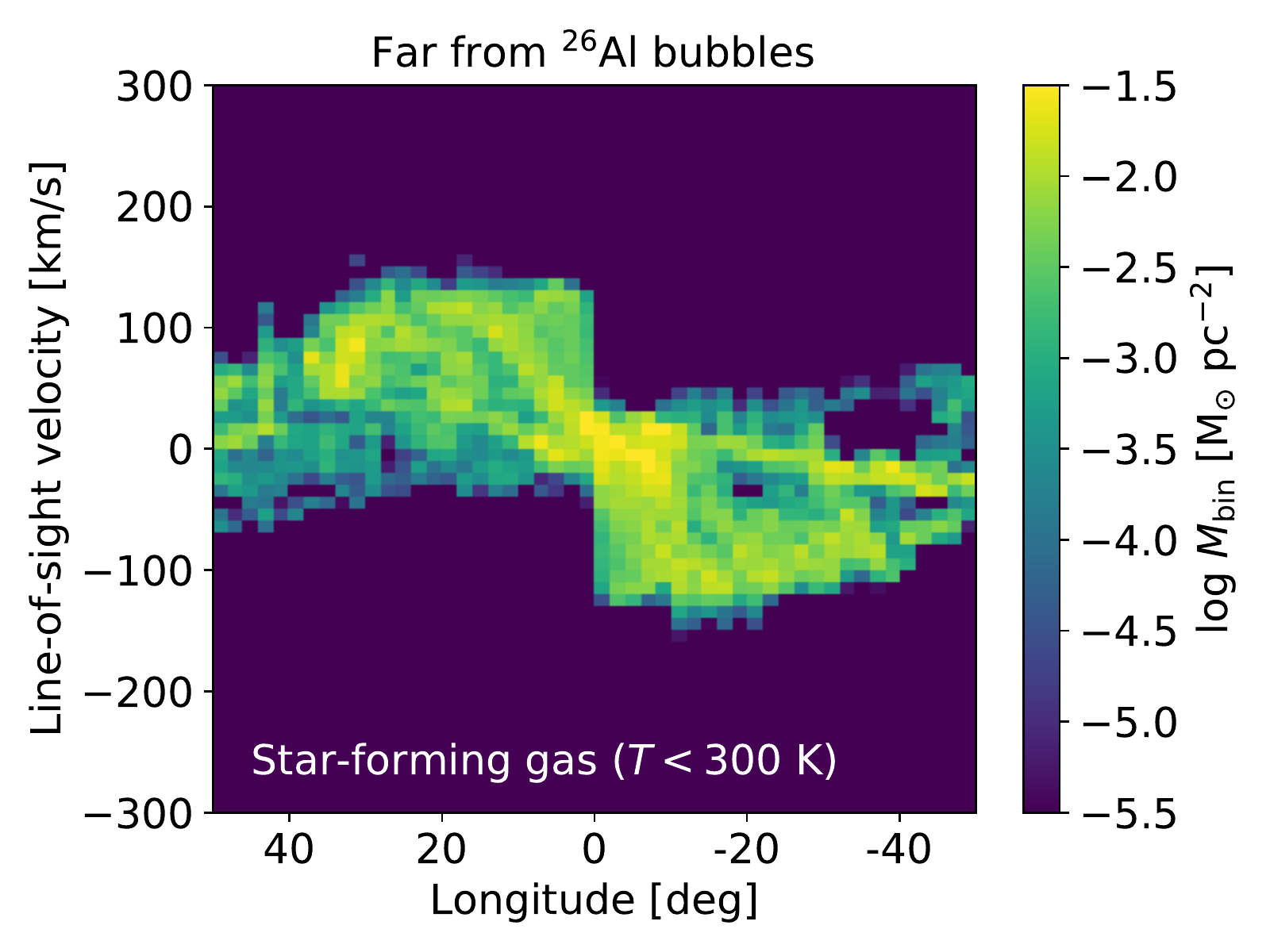}
    \caption{Mass distributions of $^{26}\mathrm{Al}$ and star-forming gas ($T < 300$ K) with respect to latitude vs. longitude and line-of-sight velocity vs. longitude, as viewed from two possible Solar positions, one inside (left column) and one far from (right column) $^{26}\mathrm{Al}$ bubbles, as shown in Fig.~\ref{fig: 26Al_projection_with_gas_contour}. The mass is weighted by its inverse squared distance relative to the observer. In the line-of-sight velocity vs. longitude plots for $^{26}\mathrm{Al}$, the $\gamma$-ray data from \citet{KretschmerEtAl2013} is overlaid (orange points with error bars).}
    \label{fig: longitude profiles}
\end{figure*}

Our alternative scenario is that the apparent large scale height and fast rotation of $^{26}\mathrm{Al}$ is a result of the Solar system being located near to or inside a large $^{26}\mathrm{Al}$ bubble, such that the $\gamma$-ray observations have been detecting $^{26}\mathrm{Al}$ emission mostly from the foreground local structures, rather than the background Galactic-scale distributions. 
To examine this scenario, in Fig.~\ref{fig: longitude profiles} we show latitude vs.~longitude ($l$-$b$) and line-of-sight velocity vs. longitude ($l$-$v$) diagrams for $^{26}\mathrm{Al}$ and star-forming gas ($T < 300$ K), using two different observer positions that we model for the Solar positions in the Galactic disc: one is located inside a $^{26}\mathrm{Al}$ bubble, and the other is located far from any bubbles, as shown in Fig.~\ref{fig: 26Al_projection_with_gas_contour}. Both of these positions lie at the Solar Circle, i.e., both are $\approx 8.0$ kpc from the Galactic centre, and we have chosen them simply as examples; we have not optimised their placement for agreement with the observations. In the case that the Sun is placed far from $^{26}\mathrm{Al}$ bubbles, the emission comes mostly from the background Galactic-scale distribution, which lacks high/low latitude material above the disc. In the case where we place the Sun inside a $^{26}\mathrm{Al}$ bubble, on the other hand, we find substantial amounts of $^{26}\mathrm{Al}$ distributed to higher/lower latitudes, and we can see the core of the $^{26}\mathrm{Al}$ bubble around $(l, b) = (20, -10)$. An analogous feature is visible in the diagram of the star-forming gas\footnote{The star-forming gas map is somewhat pixelated because the off-plane emission is coming from structures close to the observer, and, for these small distances, our 20 pc cells correspond to fairly poor angular resolution.}, and it is worth noting that similar CO features extending to $\sim 20-30^\circ$ above and below the plane, such as the Orion Complex and the Taurus-Perseus-Auriga Complex, are in fact observed in all-sky CO maps. 
The broad distribution of $^{26}\mathrm{Al}$ we find when the Sun is located at the edge of a $^{26}\text{Al}$ bubble strongly suggests that that the large scale height of $^{26}\mathrm{Al}$ estimated from $\gamma$-ray observations could be a result of contamination by foreground local structures. In this regard, our conclusion is consistent with the analysis of our previous simulation \citep{FujimotoKrumholzTachibana2018} by \citet{Pleintinger2019}, who produce simulated observations of the $^{26}\mathrm{Al}$ emission, and find that the scale height estimated from the simulated sky map is consistent with that of the observation only when the observer is placed inside a supperbubble filled with a fresh $^{26}\mathrm{Al}$.

The $l$-$v$ diagrams also support this scenario, albeit somewhat more weakly. Although we do not see systematic large velocity offsets between $^{26}\mathrm{Al}$ and star-forming gas as shown in the observations, even when we place the Sun inside a $^{26}\text{Al}$ bubble, there is a clear excess of high-velocity emission in this case, at velocities consistent with the observed values of \citet{KretschmerEtAl2013}. In addition, the observed $l$-$v$ diagram shows asymmetry about the Galactic centre, something that occurs naturally in a scenario where the observed $^{26}\mathrm{Al}$ emission comes mostly from the foreground local structures. In fact, our $l$-$v$ diagram in the case where the place the Sun inside an $^{26}\mathrm{Al}$ bubble shows an asymmetric distribution due to local structures, e.g. the regions around $(l, v) = (30, 200)$ and $(-30, -200)$. The $l$-$v$ diagrams of the star-forming gas, however, do not show any excess high velocity emission. This is again is consistent with the CO observations \citep{DameHartmannThaddeus2001}.

\section{Conclusions}
\label{sec: Conclusions}
We have performed an $N$-body+hydrodynamics simulation of a Milky-Way-type galaxy and have investigated the galactic-scale distribution and kinematics of $^{26}\mathrm{Al}$ produced in massive stars' stellar winds and SNe, including a multi-phase ISM and multi-form stellar feedback.

\begin{itemize}
    \item Transient and recurrent multi-arm spirals form naturally due to instabilities in the combined star-gas fluid, as shown in previous numerical galaxy simulations \citep[e.g.][]{WadaBabaSaitoh2011, GrandEtAl2015, BabaSaitohWada2013}. The spiral arms are short-lived and non-stationary. They continuously emerge and dissipate on timescales of $\sim$ 100 Myr, rather than forming a long-lived density wave that propagates through the galactic disc (Fig.~\ref{fig: galaxy_projections}). 
    \item Sub-kpc scale $^{26}\mathrm{Al}$ bubbles form along the gaseous spiral arms. Because the arms are co-moving with material, gas falls into them from both the leading and trailing sides of the spiral. This flow pattern is quite unlike the galactic shock induced by a density wave where the gas flows through the spiral. As a result, star clusters and the resulting $^{26}\mathrm{Al}$ bubbles form in the middle of the spiral arm, and there is no systematic preference for the direction relative to the arm in which they expand (Fig.~\ref{fig: 26Al_projection_with_gas_contour}).
    \item The scale height of $^{26}\mathrm{Al}$ is similar to that of the cold ISM, one order of magnitude smaller than the apparent observed value, even though our simulation produces supperbubbles along the spiral arms (Fig.~\ref{fig: radius_vs_height} and Fig.~\ref{fig: radius_vs_scale_height}). Similarly, the mean rotation velocity of $^{26}\mathrm{Al}$ is similar to that of the cold ISM, and 100 - 200 $\mathrm{km\ s^{-1}}$ slower than the apparent observed values (Fig.~\ref{fig: radius_vs_circular_velocity}).
    \item Our findings are inconsistent with the hypothesis that the discrepancy between the observationally-inferred scale height and rotation speed of $^{26}\text{Al}$ and that of the cold ISM is a result of preferential expansion of $^{26}\text{Al}$-rich superbubbles into the low-density regions forward of spiral arms. This hypothesis requires an asymmetry between the leading and trailing sides of spiral arms that is expected only for long-lived density wave-type arms. However, observations suggest that the Milky Way's arms are transient and material rather than long-lived and propagating. Our simulations show that arms of this type do not produce the hypothesised $^{26}\text{Al}$-asymmetry.
    \item Our findings are consistent with the hypothesis that the Solar system is located near to or inside a large $^{26}\mathrm{Al}$ bubble, and that the anomalous scale height and rotation speed occur because the $\gamma$-ray observations have been detecting $^{26}\mathrm{Al}$ emission mostly from foreground local structures, rather than background Galactic-scale distribution. In fact, when we make synthetic $l$-$b$ and $l$-$v$ diagrams of $^{26}\mathrm{Al}$ emission for a hypothetical Solar position inside a $^{26}\mathrm{Al}$ bubble, we succeed in reproducing the major qualitative features of the anomaly: a broad vertical distribution and velocity excess of $^{26}\mathrm{Al}$ as shown in $\gamma$-ray observations (Fig.~\ref{fig: longitude profiles}). This scenario is also consistent the recent discovery of live $^{60}\textrm{Fe}$ on Earth, which indicates that the Solar System has encountered core collapse SN ejecta within the last a few Myr \citep{WallnerEtAl2016, BreitschwerdtEtAl2016, SchulreichEtAl2017, KollEtAl2019}.
\end{itemize}

\section*{Acknowledgements}
YF would like to thank Junichi Baba for his useful comments on the spiral arm formation theory. MRK acknowledges support from the Australian Research Council through \textit{Future Fellowship} FT180100375. SI is supported by JSPS KAKENHI Grant Numbers 16H02160, 18H05436, and 18H05437. Simulations were carried out on the Cray XC50 at the Center for Computational Astrophysics (CfCA) of the National Astronomical Observatory of Japan and the Gadi at the National Computational Infrastructure (NCI), which is supported by the Australian Government. Computations described in this work were performed using the publicly available \textsc{enzo} code (\citealt{BryanEtAl2014}; \url{http://enzo-project.org}), which is the product of a collaborative effort of many independent scientists from numerous institutions around the world. Their commitment to open science has helped make this work possible. We acknowledge extensive use of the \textsc{yt} package (\citealt{TurkEtAl2011}; \url{http://yt-project.org}) in analysing these results and the authors would like to thank the \textsc{yt} development team for their generous help.

\section*{Data availability}
The data underlying this article will be shared on reasonable request to the corresponding author.

%%%%%%%%%%%%%%%%%%%%%%%%%%%%%%%%%%%%%%%%%%%%%%%%%%



\bibliographystyle{mnras}
\bibliography{reference} % if your bibtex file is called example.bib

%\begin{thebibliography}{99}
%\end{thebibliography}

%%%%%%%%%%%%%%%%%%%%%%%%%%%%%%%%%%%%%%%%%%%%%%%%%%

%%%%%%%%%%%%%%%%% APPENDICES %%%%%%%%%%%%%%%%%%%%%

%\appendix

%%%%%%%%%%%%%%%%%%%%%%%%%%%%%%%%%%%%%%%%%%%%%%%%%%

% Don't change these lines
\bsp	% typesetting comment
\label{lastpage}
\end{document}